\documentclass[aps,prb,preprint,showpacs,superscriptaddress,citeautoscript]{revtex4-1}

\usepackage{dcolumn}
\usepackage{bm,amsmath,amssymb,textcomp,mathrsfs}
\usepackage{graphicx}

\begin{document}

\title{
First-principles calculations of indirect Auger recombination in nitride semiconductors
}

\author{Emmanouil Kioupakis}
\affiliation{Materials Department, University of California, Santa Barbara, California 93106-5050, U.S.A.}
\affiliation{Department of Materials Science and Engineering, University of Michigan, Ann Arbor, Michigan 48109, U.S.A.}
\author{Daniel Steiauf}
\affiliation{Materials Department, University of California, Santa Barbara, California 93106-5050, U.S.A.}
\author{Patrick Rinke}
\affiliation{Materials Department, University of California, Santa Barbara, California 93106-5050, U.S.A.}
\affiliation{COMP/Department of Applied Physics, Aalto University, P.O. Box 11100, Aalto FI-00076, Finland}
\author{Kris T. Delaney}
\affiliation{Materials Department, University of California, Santa Barbara, California 93106-5050, U.S.A.}
\author{Chris G. Van de Walle}
\affiliation{Materials Department, University of California, Santa Barbara, California 93106-5050, U.S.A.}

\date{\today}

\begin{abstract}
Auger recombination is an important non-radiative carrier recombination mechanism in many classes of optoelectronic devices.
The microscopic Auger processes can be either direct or indirect, mediated by an additional scattering mechanism such as the
electron-phonon interaction and alloy disorder scattering.
Indirect Auger recombination is particularly strong in nitride
materials and affects the efficiency of nitride optoelectronic devices at high powers.
Here we present a first-principles computational formalism for the study of direct and indirect
Auger recombination in direct-band-gap semiconductors and apply it to the case of nitride materials.
We show that direct Auger recombination is weak in the nitrides
and cannot account for experimental measurements. On the other hand, carrier scattering by
phonons and alloy disorder enables indirect Auger processes that can explain the observed loss in devices.
We analyze the dominant phonon contributions to the Auger recombination rate and the influence of temperature and
strain on the values of the Auger coefficients.
Auger processes assisted by charged-defect scattering are much weaker than the phonon-assisted ones
for realistic defect densities and not important for the device performance.
The computational formalism is general and can be applied to the calculation of the Auger coefficient in other classes
of optoelectronic materials.
\end{abstract}

\maketitle


\section{Introduction}

\begin{figure}
\includegraphics[width=\columnwidth]{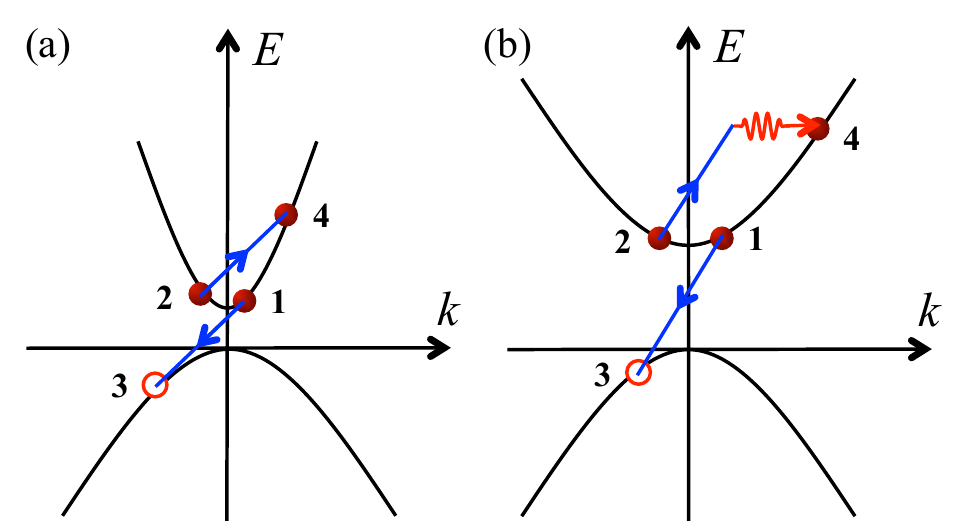}
\caption{\label{fig:auger_direct_indirect}
(Color online) Schematic diagram of (a) direct and (b) indirect electron-electron-hole Auger recombination processes.
(a) In the direct case, an electron in the conduction band (1)
recombines with a hole in the valence band (3)
while the excess energy and momentum is transferred to a second electron (2) that gets excited to a higher conduction-band state (4).
(b) The indirect process is assisted by a carrier-scattering mechanism,
such as electron-phonon coupling, alloy disorder, or defect scattering,
which provides additional momentum and
enables Auger transitions to a wider range of final conduction-band states throughout the first Brillouin zone.
}
\end{figure}

Auger recombination is an important non-radiative carrier recombination mechanism that is well recognized as a loss mechanism
in optoelectronic devices such as light-emitting (LEDs) and laser diodes\cite{Schubert_book}.
In the Auger process, the energy released through an electron-hole recombination event,
approximately equal to the band gap of the material,
is transferred via Coulomb scattering to a third free carrier
that is excited
to a higher-energy state [Fig.~\ref{fig:auger_direct_indirect}(a)].
The third carrier can be either an electron [in which case the process is called an electron-electron-hole (e-e-h) process] or a hole
 [hole-hole-electron (h-h-e)].
The overall energy and momentum of the carriers is conserved, and for non-degenerate carriers the recombination rate scales
with the third power of the carrier density.
As a result, it becomes an important carrier recombination mechanism at high carrier densities and reduces the efficiency of solid-state light emitters at high currents.
In particular, Auger recombination has been shown to be involved in the efficiency reduction
of nitride LEDs and lasers at high power
\cite{Gardner_et_al_2007,Shen_et_al_2007,Meneghini2009,Zhang2009,Laubsch2009,Laubsch2010,David2010a,David2010b,Scheibenzuber2011,Brender2011,kioupakis:231107,PhysRevLett.110.177406}.
The direct Auger recombination process [Fig.~\ref{fig:auger_direct_indirect}(a)]
has been studied both analytically (e.g., Refs. \onlinecite{BeattieLandsberg_1959,sugimura:4405,dutta:1236,PSSB:PSSB2220890204})
as well as with first-principles calculations\cite{LaksPantelides1990,PhysRevLett.89.197601,DelaneyRinkeAuger,Heinemann2011,ossicini}.

An additional type of Auger recombination mechanisms are those assisted
by scattering due to phonons, defects, alloying, etc. [Fig.~\ref{fig:auger_direct_indirect}(b)].
The coupling of charge carriers to lattice vibrations in semiconductors gives rise to important quantum processes such as
optical absorption in indirect-gap materials\cite{PhysRevLett.108.167402,Patrick2014} and free-carrier absorption in semiconductors\cite{PhysRevB.81.241201,APEX.3.082101} and
transparent conducting oxides\cite{peelaers:011914}.
Phonon-assisted Auger processes have until recently only been treated with model calculations
for the band structure and/or the electron-phonon coupling matrix elements,
and/or only a fraction of all the relevant microscopic processes have been accounted for \cite{Eagles1961,Huldt1976,Lochmann1977,Lochmann1978,Haug1977537,Pasenow2009,Bertazzi_phonon}.
We have developed a set of first-principles computational approaches to study direct and indirect Auger recombination in
nitride materials.
Some initial results of our work have been reported elsewhere.\cite{kioupakis:161107,Kioupakis2013,Steiauf2014}
Our calculations allowed us to conclude that Auger recombination, and in particular the indirect process assisted by electron-phonon coupling or
alloy-disorder scattering,
is
especially
strong in nitride materials.
The calculated coefficient is sufficient to explain the efficiency droop in nitride light emitters. Moreover, we have found that phonon-mediated processes dominate Auger recombination in GaAs\cite{Steiauf2014} and NaI.\cite{McAllister2015}

In the present paper, we present the full formalism and computational details for the
first-principles calculation of direct and indirect Auger recombination in direct-band-gap semiconductors.
Our computational formalism makes use of an array of first-principles tools, such as the maximally localized Wannier function method
for the efficient interpolation of energy eigenvalues\cite{MarzariVanderbilt97,SouzaMarzariVanderbilt01,wannier90,wannier_review}
and density functional perturbation theory\cite{Baroni_et_al_2001,QuantumEspresso} for the calculation of phonon frequencies and
electron-phonon coupling matrix elements.
We discuss the results we obtained for wurtzite nitride materials and we show that phonon- and alloy-scattering-assisted Auger recombination
dominates in this technologically important class of wide-band-gap semiconductors.
The approximations we developed to reduce the computational cost are also described.

The article is organized as follows.
In Section~\ref{sec:equations} we present the relevant equations
for the calculation of the direct and indirect Auger coefficients in semiconductors.
In Section~\ref{sec:formalism} we describe the computational formalism we implemented and the approximations we used for the calculations.
In Section~\ref{sec:results} we present our results for nitride materials:
In Subsection~\ref{sec:direct_gan} we discuss our results for direct Auger recombination in GaN.
In Subsection~\ref{sec:direct_ingan} we present our results for the alloy-scattering-assisted Auger recombination in In$_{0.25}$Ga$_{0.75}$N.
In Subsection~\ref{sec:phonon_gan}
we describe the results for the phonon-assisted Auger recombination in GaN, analyze the contributions by the various phonons,
and discuss the effects of temperature and strain.
In Subsection~\ref{sec:defect_gan} we discuss charge-defect-assisted Auger recombination in GaN, and
finally in Subsection~\ref{sec:total_auger} we present the cumulative values of the Auger recombination coefficients.

\section{Auger recombination formalism}
\label{sec:equations}
\subsection{Direct Auger recombination}
The formalism for the calculation of the direct Auger recombination rate has been reported previously\cite{Ridley,LaksPantelides1990,PhysRevLett.89.197601,ossicini}; for completeness we provide a brief review here.
The Auger recombination rate is calculated starting from Fermi's golden rule, which
gives the transition probability per unit time in terms of matrix elements of the perturbation and the quasiparticle energies.
For the case of electrons and holes in a solid, the Auger recombination rate is given by:
\begin{align}{\label{eq:auger_rate}}
R = 2 \frac{2\pi}{\hbar} \sum_{\bm{1234}} P |M_{\bm{1234}}|^2 \delta(\epsilon_{\bm{1}}+\epsilon_{\bm{2}}-\epsilon_{\bm{3}}-\epsilon_{\bm{4}}),
\end{align}
where the bold indices are composite band and $\bm{k}$-point indices [$\bm{1}\equiv (n_1,\bm{k}_1)$],
the factor of 2 accounts for spin,
$P$ is a statistics factor that accounts for the occupation numbers and ensures that transitions occur only from
occupied to empty fermion states,
\[ P=f_{\bm{1}}f_{\bm{2}}(1-f_{\bm{3}})(1-f_{\bm{4}}), \]
and $f$ are free-carrier occupation numbers according to Fermi-Dirac statistics.
The $\delta$-function ensures energy conservation, while momentum conservation is imposed by the matrix elements.

\begin{figure}
\includegraphics[width=\columnwidth]{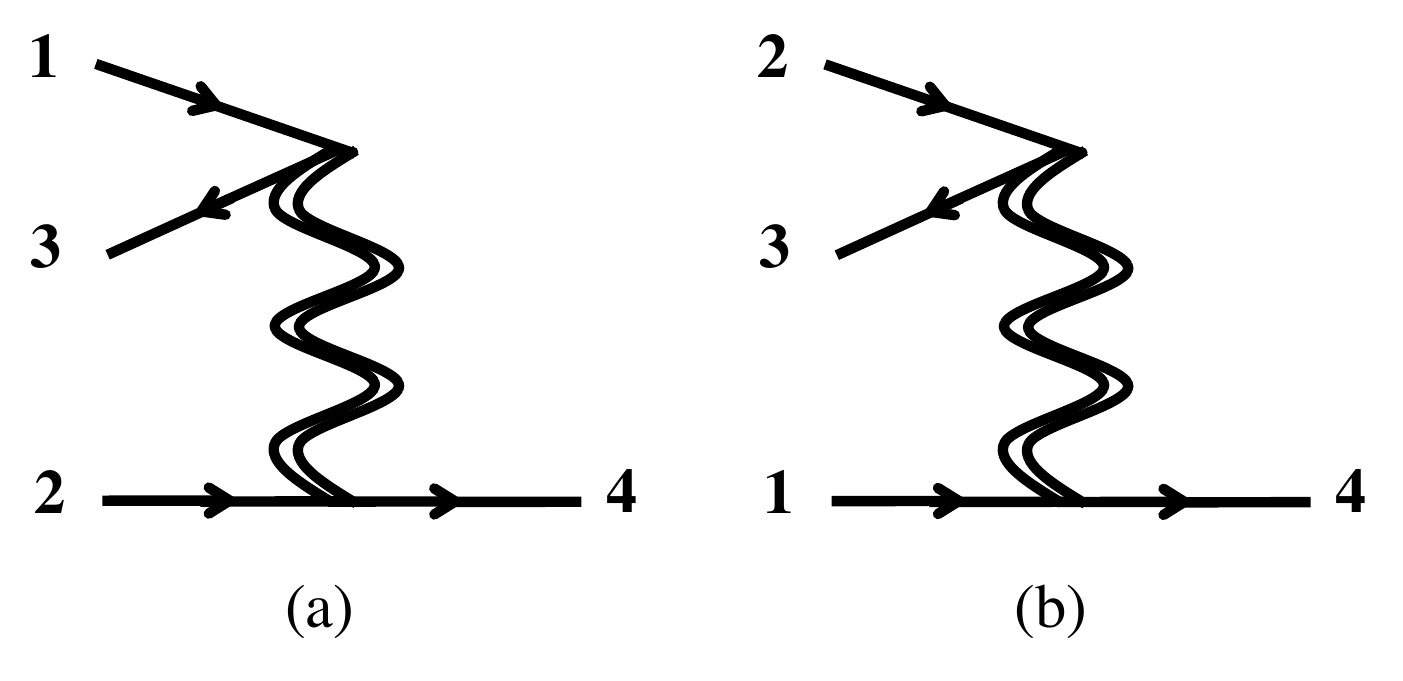}
\caption{\label{fig:auger_direct_diagrams} Feyman diagrams for the (a) direct and (b) exchange terms of the Auger recombination process of Eq.(~\ref{eq:direct_auger_ME}). Diagram (a) corresponds to the direct process shown in Fig.~\ref{fig:auger_direct_indirect}(a).}
\end{figure}

The perturbing Hamiltonian is the screened Coulomb interaction between the carriers and involves two terms,
the direct and the exchange one, that account for the antisymmetry of the many-body wave function under fermion exchange
\cite{Ridley}:
\begin{equation}\label{eq:direct_auger_ME}
|M_{\bm{1234}}|^2 \equiv |M^d_{\bm{1234}}-M^x_{\bm{1234}}|^2 + |M^d_{\bm{1234}}|^2 +|M^x_{\bm{1234}}|^2,
\end{equation}
where the direct ($M^d$) and exchange ($M^x$) terms are given by matrix elements of the screened Coulomb interaction ($W$) between electron and hole wave functions ($\psi$):
\begin{align}
M^d_{\bm{1234}} & \equiv \langle \psi_{\bm{1}} \psi_{\bm{2}} | W | \psi_{\bm{3}} \psi_{\bm{4}} \rangle, \\
M^x_{\bm{1234}} & \equiv \langle \psi_{\bm{1}} \psi_{\bm{2}} | W | \psi_{\bm{4}} \psi_{\bm{3}} \rangle,
\end{align}
and are shown in Figs.~\ref{fig:auger_direct_diagrams}(a) and (b), respectively.
The matrix elements of the screened Coulomb interaction are given by:
\begin{eqnarray}{\label{eq:auger_me}}
\lefteqn{\langle \psi_{\bm{1}} \psi_{\bm{2}} | W | \psi_{\bm{3}} \psi_{\bm{4}} \rangle  =}  \nonumber \\
&  = &\iint d\bm{r}_1 d\bm{r}_2 \psi_{\bm{1}}^*(\bm{r}_1)\psi_{\bm{2}}^*(\bm{r}_2)W(\bm{r}_1,\bm{r}_2) \psi_{\bm{3}}(\bm{r}_1)\psi_{\bm{4}}(\bm{r}_2)  \nonumber \\
& = &\frac{1}{V} \sum_{\bm{G}} \delta_{\bm{k}_1+\bm{k}_2,\bm{k}_3+\bm{k}_4+\bm{G}'} \tilde{W}(\bm{k}_1-\bm{k}_3+\bm{G}) \nonumber \\
& & \times I_{{\bm{1}},{\bm{3}}}(\bm{G})I_{{\bm{2}},{\bm{4}}}(\bm{G}'-\bm{G}),
\end{eqnarray}
where $\bm{G}'$ is the Umklapp vector that brings $\bm{k}_4=\bm{k}_1+\bm{k}_2-\bm{k}_3+\bm{G}'$ back into the first Brillouin zone.
Ignoring local-field effects\cite{HybertsenLouie86}, the Fourier-transformed Coulomb interaction ($\tilde{W}$) is given by:
\begin{align}{\label{eq:Wq}}
W(\bm{r}_1,\bm{r}_2) & = W(\bm{r}_1-\bm{r}_2)\equiv \frac{1}{V}\sum_{\bm{q}} \tilde{W}(\bm{q})e^{i\bm{q}\cdot(\bm{r}_1-\bm{r}_2)} \nonumber \\
&=\frac{1}{V}\sum_{\bm{q}}\frac{1}{\varepsilon(\bm{q})} \frac{4\pi e^2}{q^2+\lambda^2}e^{i\bm{q}\cdot(\bm{r}_1-\bm{r}_2)}
\end{align}
where $\varepsilon(\bm{q})$ is the static dielectric function of the material and $\lambda$ is the inverse screening length
due to the free carriers.
The overlap integrals, $I$, are given by:
\begin{align}{\label{eq:I}}
I_{{\bm{\alpha}},{\bm{\beta}}}(\bm{G}) & \equiv \sum_{\bm{G}_1} c_{{\bm{\alpha}}}^*(\bm{G}_1)c_{{\bm{\beta}}}(\bm{G}_1-\bm{G}) \nonumber \\
 & =\int_{{\text{cell}}} u_{{\bm{\alpha}}}^*(\bm{r})u_{{\bm{\beta}}}(\bm{r})e^{i\bm{G}\cdot\bm{r}}d\bm{r},
\end{align}
where $c$ are the plane-wave components of the lattice-periodic part of the Bloch functions $u$.

The Auger rate is then calculated using:
\begin{align}{\label{eq:auger_direct}}
\frac{dn}{dt} & =\frac{R}{V} = \frac{4\pi}{\hbar} \frac{1}{V_{\text{cell}}^3} \frac{1}{N_{\bm{k}}^3}\sum_{\bm{123}n_4}f_{\bm{1}}f_{\bm{2}}(1-f_{\bm{3}})(1-f_{\bm{4}}) \nonumber \\
& \qquad \times |VM_{\bm{1234}}|^2 \delta( \epsilon_{\bm{1}} +\epsilon_{\bm{2}}-\epsilon_{\bm{3}}-\epsilon_{\bm{4}}),
\end{align}
where $\bm{4}\equiv(n_4,\bm{k}_1+\bm{k}_2-\bm{k}_3+\bm{G}')$.
In the remainder of this paper, we assume an equal density of free electrons and holes ($n=p$),
which is usually the case in optoelectronic devices for high injected carrier densities and under steady-state conditions.
For non-degenerate carriers, which can be described with Boltzmann statistics, the Auger recombination rate
scales with the third power of the carrier density\cite{Ridley}.
At high carrier concentrations, however, the carriers become degenerate and need to be described by Fermi-Dirac statistics.
In this case, the exponent of the power law can take values lower than 3, an effect known as phase-space filling\cite{Hader2005}.
In the following, we will calculate density-dependent Auger coefficients $C=C(n)$ defined by:
\[ C(n) \equiv \frac{R(n)}{n^3V}. \]

\subsection{Indirect Auger recombination}
In this Subsection, we present the complete formalism for the calculation of the indirect recombination rate.
Various aspects of this formalism have been presented before in Refs. \onlinecite{Eagles1961,Huldt1976,Lochmann1977,Lochmann1978,Haug1977537,Pasenow2009}.
The transition probability per unit time ($\mathscr{P}$) for indirect Auger processes from an initial state ($\text{I}$) to a final state ($\text{F}$) via an intermediate state ($\text{M}$) is determined based on second-order Fermi's
golden rule\cite{BassaniParravicini}:
\[ \mathscr{P}_{\text{I}\rightarrow\text{F}}=\frac{2\pi}{\hbar}\left|\sum_{\text{M}}\frac{\mathscr{H}_\text{IM}\mathscr{H}_{\text{MF}}}{E_{\text{M}}-E_{\text{I}}} \right|^2 \delta(E_{\text{F}}-E_{\text{I}})\]
where the perturbation Hamiltonian ($\mathscr{H}$) is the combination of the screened Coulomb interaction
plus the scattering mechanism (e.g., the electron-phonon interaction), and $E_{\text{I}}$, $E_{\text{F}}$, and $E_{\text{M}}$ are the energies of the initial, final, and intermediate states.

By summing over all initial and final states, we obtain an expression for
the recombination rate similar to the expression for direct Auger recombination [Eq.~(\ref{eq:auger_rate})].
The equation for phonon-assisted Auger recombination is:
\begin{align}{\label{eq:indirect_auger_rate}}
R = 2 \frac{2\pi}{\hbar} \sum_{\bm{1234},\nu\bm{q}} \tilde{P} |\tilde{M}_{\bm{1234};\nu\bm{q}}|^2 \delta(\epsilon_{\bm{1}}+\epsilon_{\bm{2}}-\epsilon_{\bm{3}}-\epsilon_{\bm{4}} \mp \hbar\omega_{\nu\bm{q}}),
\end{align}
where $\nu$ is the phonon mode and $\bm{q}$ is the phonon wave vector.
The upper (lower) sign corresponds to the phonon-emission (absorption) process, respectively.
The difference with direct Auger recombination is that the phonon energy ($\hbar\omega_{\nu\bm{q}}$)
enters the energy-conserving $\delta$-function (the phonon momentum $\hbar\bm{q}$ appears
in the momentum-conserving matrix elements later).
The statistics factor $P$ is modified to account for the additional phonon absorption or emission process,
\[ \tilde{P} = f_{\bm{1}}f_{\bm{2}}(1-f_{\bm{3}})(1-f_{\bm{4}})\left( n_{\nu\bm{q}} + \frac{1}{2} \pm \frac{1}{2} \right), \]
where $n_{\nu\bm{q}}$ are the phonon occupation numbers given by Bose-Einstein statistics:
\[ n_{\nu\bm{q}} = \frac{1}{e^{\hbar\omega_{\nu\bm{q}}/k_\text{B}T} - 1 }, \]
and the matrix element has been generalized to include the scattering to the virtual intermediate state.

\begin{figure}
\includegraphics[width=\columnwidth]{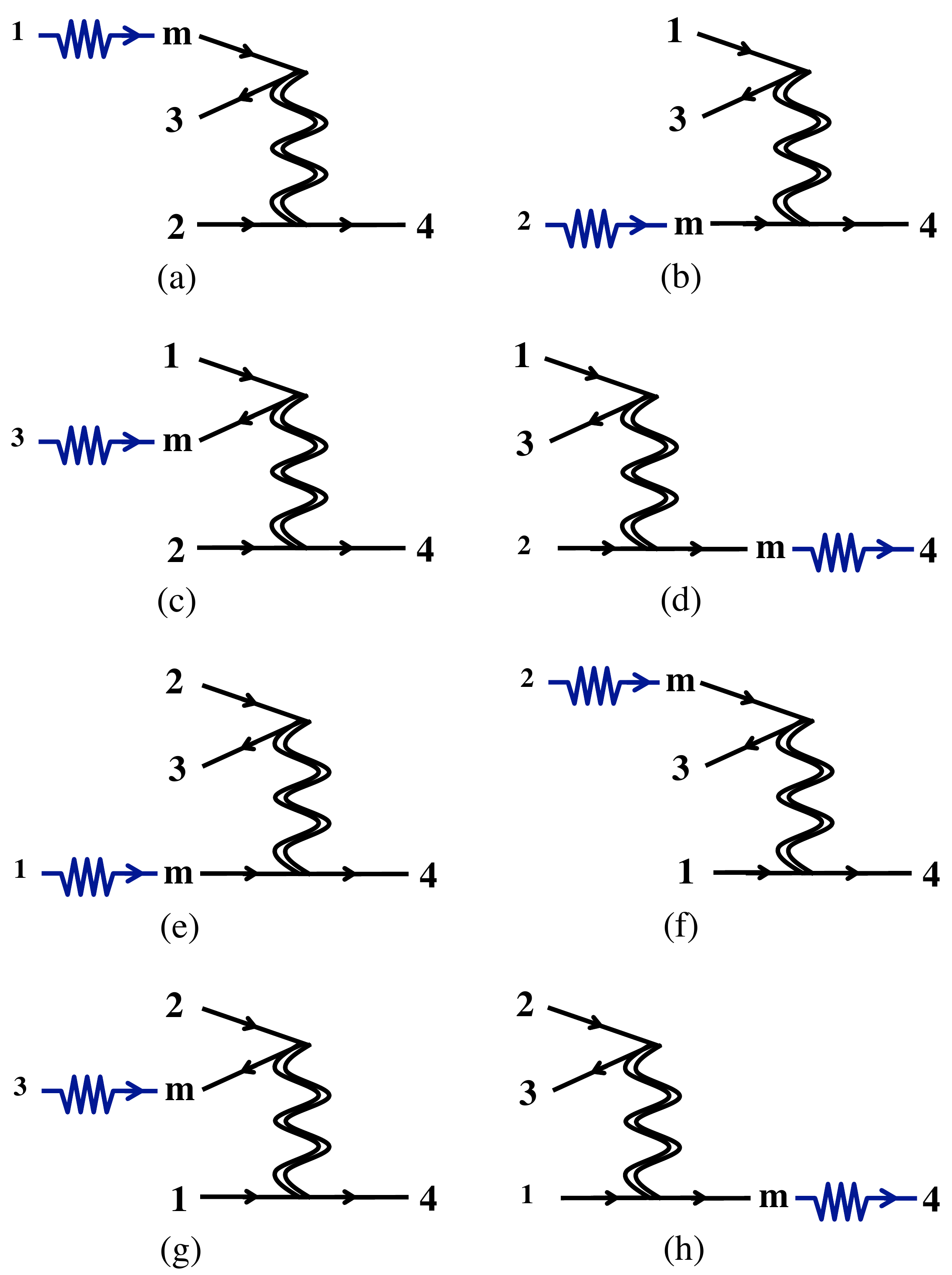}
\caption{\label{fig:auger_indirect_diagrams} (Color online) (a)--(h) Schematic diagrams corresponding to each of the microscopic indirect
Auger recombination processes of Eq.~(\ref{eq:auger_gme}) ($\tilde{M}^1$--$\tilde{M}^8$, respectively).}
\end{figure}

There are eight diagrams that enter the calculation of the generalized matrix element that correspond to all possible
combinations of the electron-phonon and screened-Coulomb interaction Hamiltonians (direct and exchange)
for both parallel and antiparallel initial carrier spin configurations.
These diagrams are summed in groups and then squared,
depending on whether they correspond to distinguishable or indistinguishable quantum processes.
The generalized matrix element ($\tilde{M}$) is given by:
\begin{align}{\label{eq:auger_gme}}
|\tilde{M}|^2 = & \left|\tilde{M}^1 +\tilde{M}^2 +\tilde{M}^3 +\tilde{M}^4 \right. \nonumber \\
& \left. -\tilde{M}^5 -\tilde{M}^6 -\tilde{M}^7 -\tilde{M}^8 \right|^2 + \nonumber \\
+ & \left| \tilde{M}^1 +\tilde{M}^2 +\tilde{M}^3 +\tilde{M}^4 \right|^2 + \nonumber \\
+ & \left| \tilde{M}^5 +\tilde{M}^6 +\tilde{M}^7 +\tilde{M}^8 \right|^2 ,
\end{align}
where each of the $\tilde{M}^1$ to $\tilde{M}^8$ terms corresponds to each of the phonon-assisted process depicted in Figs.~\ref{fig:auger_indirect_diagrams}(a)--(h):
\begin{align}\label{eq:auger_gme_1_8}
\tilde{M}^1_{\bm{1234};\nu\bm{q}} & = \sum_{\bm{m}} \frac{g_{\bm{1}\bm{m};\nu}M^d_{\bm{m234}}}{\epsilon_{\bm{m}}-\epsilon_{\bm{1}} \pm \hbar\omega_{\nu\bm{q}} +i\eta} \\
\tilde{M}^2_{\bm{1234};\nu\bm{q}} & = \sum_{\bm{m}} \frac{g_{\bm{2}\bm{m};\nu}M^d_{\bm{1m34}}}{\epsilon_{\bm{m}}-\epsilon_{\bm{2}} \pm \hbar\omega_{\nu\bm{q}} +i\eta} \\
\tilde{M}^3_{\bm{1234};\nu\bm{q}} & = \sum_{\bm{m}} \frac{M^d_{\bm{12m4}}g_{\bm{m}\bm{3};\nu}}{\epsilon_{\bm{m}}-\epsilon_{\bm{3}} \mp \hbar\omega_{\nu\bm{q}} +i\eta} \\
\tilde{M}^4_{\bm{1234};\nu\bm{q}} & = \sum_{\bm{m}} \frac{M^d_{\bm{123m}}g_{\bm{m}\bm{4};\nu}}{\epsilon_{\bm{m}}-\epsilon_{\bm{4}} \mp \hbar\omega_{\nu\bm{q}} +i\eta} \\
\tilde{M}^5_{\bm{1234};\nu\bm{q}} & = \sum_{\bm{m}} \frac{g_{\bm{1}\bm{m};\nu}M^x_{\bm{m234}}}{\epsilon_{\bm{m}}-\epsilon_{\bm{1}} \pm \hbar\omega_{\nu\bm{q}} +i\eta} \\
\tilde{M}^6_{\bm{1234};\nu\bm{q}} & = \sum_{\bm{m}} \frac{g_{\bm{2}\bm{m};\nu}M^x_{\bm{1m34}}}{\epsilon_{\bm{m}}-\epsilon_{\bm{2}} \pm \hbar\omega_{\nu\bm{q}} +i\eta} \\
\tilde{M}^7_{\bm{1234};\nu\bm{q}} & = \sum_{\bm{m}} \frac{M^x_{\bm{12m4}}g_{\bm{m}\bm{3};\nu}}{\epsilon_{\bm{m}}-\epsilon_{\bm{3}} \mp \hbar\omega_{\nu\bm{q}} +i\eta} \\
\tilde{M}^8_{\bm{1234};\nu\bm{q}} & = \sum_{\bm{m}} \frac{M^x_{\bm{123m}}g_{\bm{m}\bm{4};\nu}}{\epsilon_{\bm{m}}-\epsilon_{\bm{4}} \mp \hbar\omega_{\nu\bm{q}} +i\eta},
\end{align}
and $\eta$ is the inverse of the lifetime of the intermediate state. For example, the process in Fig.~\ref{fig:auger_direct_indirect}(b) is given by the $\tilde{M}^4$ term [Fig.~\ref{fig:auger_indirect_diagrams}(d)].
We made sure to evaluate all terms using the same set of calculated wave functions in order
to preserve the correct phase information and ensure that the cross terms among the various paths are properly taken into account.

The electron-phonon coupling matrix elements ($g$) are given in terms of the electron and hole wave functions and the derivative $\partial_{\nu\bm{q}}V$ of the self-consistent potential due to a collective ionic displacement by phonon mode $\nu\bm{q}$\cite{PhysRevB.76.165108}:
\[ g_{n\bm{k},m\bm{k}+\bm{q};\nu} = \left( \frac{\hbar}{2M_0\omega_{\nu\bm{q}}} \right)^{1/2}\langle \psi_{n\bm{k}} | (\partial_{\nu\bm{q}}V)^* |\psi_{m\bm{k}+\bm{q}} \rangle,  \]
where $M_0$ is the total mass of the atoms in the unit cell.

\section{Computational formalism}
\label{sec:formalism}
Calculations for the electronic energies and wave functions
were performed using density functional theory (DFT)\cite{HohenbergKohn64,KohnSham65}
within the local-density approximation for the exchange-correlation potential\cite{CeperleyAlder80,PerdewZunger81}
and the plane-wave pseudopotential method\cite{IhmZungerCohen79,QuantumEspresso}.
The semicore 3\emph{d} electrons of Ga and 4\emph{d} electrons of In are treated as valence electrons
to get accurate electronic and structural properties\cite{Stampfl1999a}.
Effects due to the spin-orbit interaction, which are small in nitride semiconductors, were not included in the calculations.
After a self-consistent calculation was performed to obtain the charge density, we determined the maximally-localized Wannier
functions\cite{MarzariVanderbilt97,SouzaMarzariVanderbilt01,wannier90} which we subsequently used
to interpolate the band structure from coarse $\bm{k}$-grids (8$\times$8$\times$8 for the GaN calculations in Sec.~\ref{sec:direct_gan} and 4$\times$4$\times$4 for the In$_{0.25}$Ga$_{0.75}$N calculations described in Sec.~\ref{sec:direct_ingan})
to arbitrarily fine $\bm{k}$-grids (as fine as 64$\times$64$\times$32 for GaN and 24$\times$24$\times$12 for In$_{0.25}$Ga$_{0.75}$N) in the first Brillouin zone.
This enables the efficient determination of the band energies near the band extrema, which we subsequently use to determine the free-carrier quasi-Fermi levels,
without the need to perform costly non-self-consistent DFT calculations for these fine $\bm{k}$-grids.
The band structure is not significantly affected by the addition of the free carriers for carrier densities
that are relevant for most devices (up to 10$^{20}$ cm$^{-3}$).
The delta functions in the rate equations are approximated by finite-width Gaussians.
All $\bm{k}$ grids mentioned in this paper are unshifted (i.e., they include the $\Gamma$ point). We expect that shifted $\bm{k}$ grids yield similar results provided they are sufficiently converged.

To describe the distribution of the free carriers over states near the band extrema,
we interpolated the bands to fine $\bm{k}$-point meshes in the Brillouin zone
and determined the quasi-Fermi levels as a function of the free-carrier density and temperature.
We then applied a cutoff criterion to limit the number of $\bm{k}$-points used
in subsequent Auger rate calculations. The criterion was to keep only $\bm{k}$-points which correspond to
band energies within an energy cutoff from the corresponding band edge,
$E_{\text{cutoff}}=\epsilon_{\text{F}}+MkT$, where $\epsilon_{\text{F}}$ is the free-carrier quasi-Fermi level
and $M$ is an integer. The integer cutoff parameter $M$ is determined such that the free-carrier density obtained
after applying the cutoff criterion
differs by less than 1\% from its starting value.
This ensures that most of the free charge
carriers are accounted for when calculating the Auger rates.
We repeat this procedure to converge the Auger coefficients with respect to the $\bm{k}$-grid spacing.

To further reduce the number of $\bm{k}$-points used in Auger rate calculations, the sums over hole states are sampled with coarser grids that are
half as dense along each linear dimension as the corresponding electron sums.
This choice relies on the fact that holes in the nitrides have a larger effective mass than electrons and thus occupy a larger $\bm{k}$-space volume around $\Gamma$
that can be effectively sampled with coarser grids.
Since we eventually converge our calculations with respect to the grid spacing, the choice of sparser hole grids speeds up the calculations without loss of accuracy.

For the calculation of the matrix elements of the screened Coulomb interaction
[Eq.~(\ref{eq:auger_me})] we used a model dielectric function\cite{Cappellini1993}:
\[ \varepsilon(q) = 1 + \{ [\varepsilon_{\infty}-1]^{-1}+\alpha (q/q_{\text{TF}})^2 + \hbar^2q^4/(4m^2\omega_p^2)\}^{-1}, \]
where $\varepsilon_{\infty}$ is the dielectric constant due to electronic screening,
$q_{\text{TF}}$ is the Thomas-Fermi wave vector,
$\omega_p$ is the plasma frequency, and $\alpha=1.563$ is an empirical parameter.
This model has been shown to accurately reproduce the first-principles dielectric function of zinc-blende GaN calculated
in the random-phase-approximation.\cite{Palummo1995393}
For the wave functions used to calculate the matrix elements we employed a lower plane-wave cut-off energy (50 Ry) than the
self-consistent field calculations (90 Ry) to speed up the calculation.
The choice of this lower plane-wave cut-off energy
affects the calculated Auger rates by at most 10\%.
For the dielectric constant of GaN needed by the model, we used the directionally averaged
experimental dielectric constant,
$\varepsilon_{\infty}=(\varepsilon_{\infty \parallel}\varepsilon_{\infty \perp}^2)^{1/3}=5.5$\cite{Manchon19701227,BarkerIlegems1973},
where $\varepsilon_{\parallel}$ and $\varepsilon_{\perp}$ are the dielectric constants for polarization parallel and perpendicular
to the $c$-axis of the wurtzite structure, respectively.
The dielectric constant of In$_{0.25}$Ga$_{0.75}$N
($\varepsilon_{\infty}=7.0$)
was estimated with a model extrapolation\cite{bergmann:1196}
to a band gap of 2.4 eV, the value of the 25\% InGaN alloy as
predicted from hybrid-density-functional calculations\cite{Moses2010}.

For the free-carrier screening wave vector of electrons $\lambda_{\text{e}}$ we used the Debye-H\"uckel equation for non-degenerate carriers
($\lambda_{\text{e}}^2=4\pi n e^2 / \epsilon_{\infty} k_{\text{B}}T$) if the quasi-Fermi energy of electrons referenced to the energy of the conduction-band minimum ($E_{\text{CBM}}$) is $\epsilon_{\text{F}} -E_{\text{CBM}}< 0$
or $0 \le \epsilon_{\text{F}} -E_{\text{CBM}}< \frac{3}{2}k_{\text{B}}T$,
and the Thomas-Fermi equation for degenerate carriers [$\lambda_{\text{e}}^2=6\pi n e^2 / \epsilon_{\infty} (\epsilon_{\text{F}} -E_{\text{CBM}})$]
if $\epsilon_{\text{F}} -E_{\text{CBM}} \ge  \frac{3}{2}k_{\text{B}}T$.
We use similar equations for the screening wave vector of holes $\lambda_{\text{h}}$ and sum the contributions by electrons and holes to obtain the total free-carrier screening wave vector $\lambda^2 = \lambda_{\text{e}}^2+\lambda_{\text{h}}^2$.

For the calculation of the phonon-assisted Auger coefficients, we need to make approximations in the way we
calculate the indirect Auger rate.
The technical challenge is that the full calculation of Eq.~(\ref{eq:indirect_auger_rate}) that accounts for
all possible initial and final electronic states is computationally very expensive.
The high computational cost arises because of the fine grids needed to sample the free-carrier distributions in reciprocal space and
the large number of associated phonon calculations.
The approximation we make is based on the fact that the free carriers are confined in reciprocal space
to the vicinity of the band extrema near the $\Gamma$ point of the Brillouin zone.
As a result, the wave functions of the initial electron and hole states can be effectively
approximated by the conduction- and valence-band wave functions at the Brillouin-zone center.
We will later show that this is a valid approximation for the case of wide-band-gap nitrides.

Within the approximation that the initial-state wave functions are those at the $\Gamma$ point, the sums over indices
$\bm{1}$ to $\bm{3}$ yield $N/2$, where $N$ is the total number of free electrons or holes and the factor of 2 arises
because we
already accounted for the carrier spin in the matrix elements. State $\bm{4}$ is initially empty
since it is located at an energy on the order of the band gap (i.e., several eV) away from the corresponding band edge.
In the case of the e--e--h process, for example, the expression we need to calculate reduces to the form:
\begin{align}{\label{eq:indirect_auger_rate_Gamma}}
R = 2 \frac{2\pi}{\hbar} \frac{N^3}{8}\sum_{n\nu\bm{q}} & \left( n_{\nu\bm{q}} +\frac{1}{2} \pm\frac{1}{2} \right) |\tilde{M}_{c\bm{\Gamma},c\bm{\Gamma},v\bm{\Gamma},n\bm{q};\nu\bm{q}}|^2 \nonumber \\
& \times \delta(\epsilon_{c\bm{\Gamma}}+\epsilon_{c\bm{\Gamma}}-\epsilon_{v\bm{\Gamma}}-\epsilon_{n\bm{q}} \mp \hbar\omega_{\nu\bm{q}}),
\end{align}
which involves only a single sum over the phonon wave vectors and is computationally feasible.
A similar expression can be derived for the h--h--e case:
\begin{align}{\label{eq:indirect_auger_rate_hhe_Gamma}}
R = 2 \frac{2\pi}{\hbar} \frac{N^3}{8}\sum_{n\nu\bm{q}} & \left( n_{\nu\bm{q}} +\frac{1}{2} \pm\frac{1}{2} \right) |\tilde{M}_{v_1\bm{\Gamma},v_2\bm{\Gamma},c\bm{\Gamma},n\bm{q};\nu\bm{q}}|^2 \nonumber \\
& \times \delta(-\epsilon_{v_1\bm{\Gamma}}-\epsilon_{v_2\bm{\Gamma}}+\epsilon_{c\bm{\Gamma}}+\epsilon_{n\bm{q}} \mp \hbar\omega_{\nu\bm{q}}),
\end{align}
where $v_1$ and $v_2$ are hole band indices.
The eight components of the
generalized matrix element in Eq.~(\ref{eq:indirect_auger_rate_Gamma}) are given by:
\begin{align}{\label{eq:gme_Gamma}}
\tilde{M}^1_{c\bm{\Gamma},c\bm{\Gamma},v\bm{\Gamma},n\bm{q};\nu\bm{q}}  & = \sum_{m} \frac{g_{c\bm{\Gamma},m\bm{q};\nu}M^d_{m\bm{q},c\bm{\Gamma};v\bm{\Gamma},n\bm{q}}}{\epsilon_{m\bm{q}}-\epsilon_{c\bm{\Gamma}} \pm \hbar\omega_{\nu\bm{q}} +i\eta}.
\end{align}
Similar expressions hold for $\tilde{M}^2$ to $\tilde{M}^8$.
Since we are neglecting the spin-orbit interaction, the heavy- and light-hole bands are degenerate at $\Gamma$,
and we average
their contribution
in the evaluation of
Eqs.~(\ref{eq:indirect_auger_rate_Gamma}) and~(\ref{eq:indirect_auger_rate_hhe_Gamma}).
The phonon frequencies and electron-phonon coupling matrix elements
were calculated using density functional perturbation theory\cite{Baroni_et_al_2001,QuantumEspresso} for $\bm{k}$-points on a
24$\times$24$\times$12 grid that fall in the irreducible part of the first Brillouin zone of GaN.


\section{Results and discussion}
\label{sec:results}
\subsection{Direct Auger recombination in GaN}
\label{sec:direct_gan}

\begin{figure}
\includegraphics[width=\columnwidth]{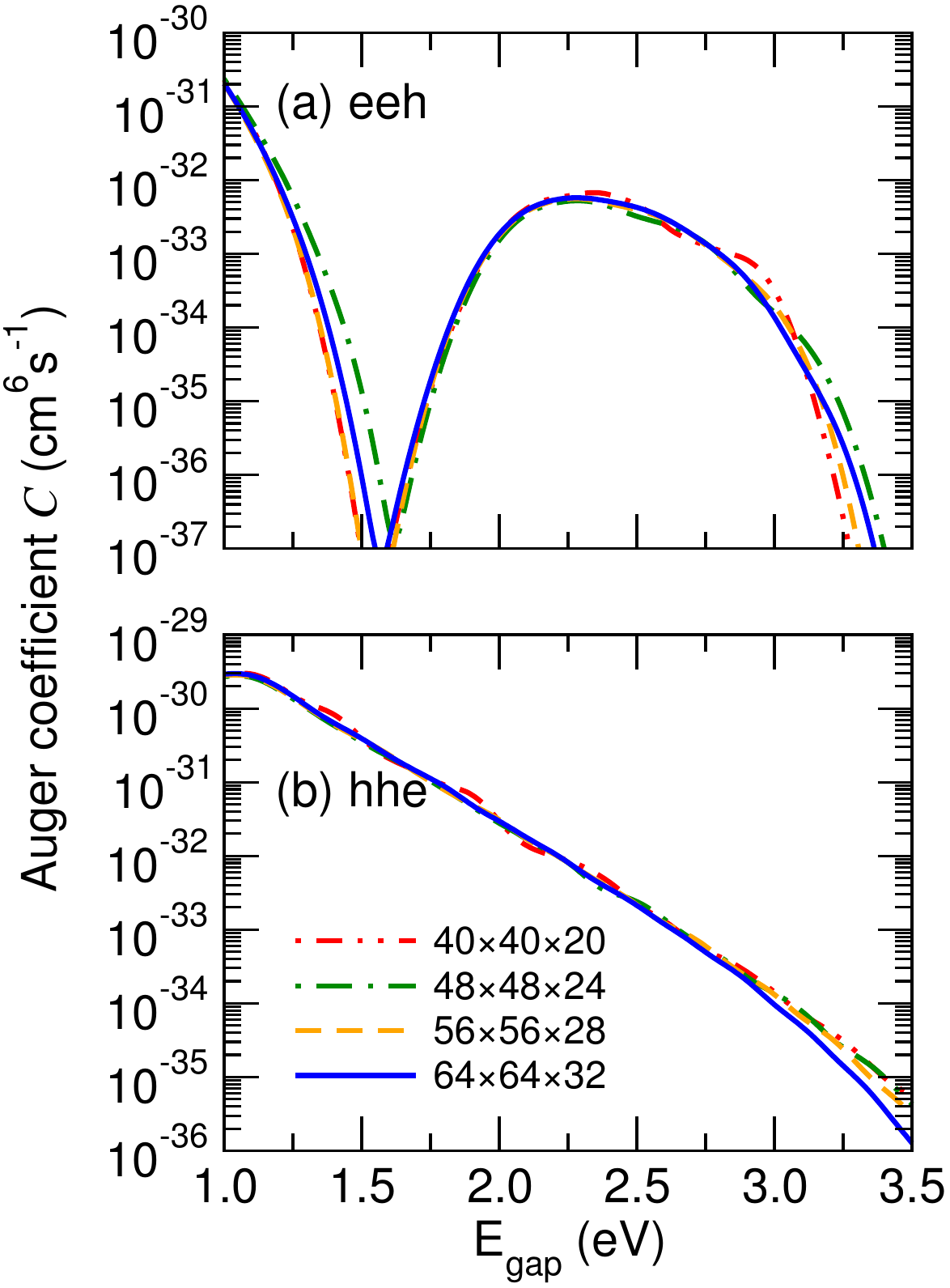}
\caption{\label{fig:gan_auger_vs_nk}
(Color online) Auger coefficients of GaN for a carrier density of 10$^{19}$ cm$^{-3}$ and a temperature of 300 K
due to (a) electron-electron-hole (e-e-h) and (b) hole-hole-electron (h-h-e) Auger
processes plotted as a function of the scissors-shift-adjusted band gap for various k-grid sampling densities.}
\end{figure}

We performed calculations for the direct Auger coefficient in bulk GaN and show that it cannot account
for the efficiency reduction of nitride LEDs.
In our calculations of Auger rates, we treat the band gap of GaN as an adjustable parameter and rigidly shift the conduction bands with a scissors operator.
This serves to correct the band-gap problem of density functional theory and to model the effect of alloying GaN with InN to form In$_x$Ga$_{1-x}$N alloys.
By varying the band-gap value, the third carrier that participates in the Auger process
is excited to a different range of higher-energy states for different gap values and this
leads to a variation of the Auger rate with respect to the adjusted band gap.
The screened-Coulomb-interaction matrix elements are calculated using the wave functions of GaN.
The Auger coefficients due to direct e--e--h and h--h--e Auger recombination processes
are plotted as a function of the adjusted band gap
in Figs.~\ref{fig:gan_auger_vs_nk}(a) and~\ref{fig:gan_auger_vs_nk}(b), respectively.
The values of the direct e--e--h and h--h--e Auger coefficients for pure GaN are simply given by the value of the Auger coefficients
for the experimental band-gap value of GaN (3.5 eV \cite{vurgaftman:3675}).
The broadening of the delta function in Eq.~(\ref{eq:auger_rate}) is set to 0.1 eV
and the free-carrier densities and temperatures to $n=p=10^{19}$ cm$^{-3}$ and 300 K, respectively.
The different curves correspond to different grids that we employed to sample the part of the
Brillouin zone occupied by free carriers, illustrating the convergence with respect to the grid spacing.

The data indicate that both e--e--h and h--h--e direct Auger processes are very weak for GaN
(less than 10$^{-35}$ cm$^6$s$^{-1}$ for $E_{\text{gap}}=3.5$ eV)
and the corresponding coefficients cannot account for the experimentally measured values
in nitride devices (10$^{-31}$--10$^{-30}$ cm$^6$s$^{-1}$).\cite{Shen_et_al_2007,Meneghini2009,Zhang2009,Laubsch2009,Laubsch2010,David2010a,David2010b,Scheibenzuber2011,Brender2011}
The coefficients increase drastically for decreasing band-gap values in the range of 2.4--3.5 eV,
which is the typical band-gap range for nitride devices.
However, even the larger Auger coefficients at lower band-gap values cannot explain the experimental measurements.
The coefficient of the h--h--e process increases exponentially for decreasing adjusted band gap and reaches a value of
3.5$\times$10$^{-33}$ cm$^6$s$^{-1}$ for band-gap values in the green ($E_{\text{gap}}=2.4$ eV).
The e--e--h process increases for decreasing band gap and displays a peak for a band-gap value of 2.3 eV due to interband Auger transitions to the
second conduction band.
The magnitude of the Auger coefficient at this peak (5.7$\times$10$^{-33}$ cm$^6$s$^{-1}$), however, is too small to account for the experimental observations.
This result is different from earlier calculations\cite{DelaneyRinkeAuger} that used a different approach
to model the Auger process.
The discrepancy has been traced back to the inadvertent omission of a normalization factor and the treatment of the long-range part of the Coulomb interaction in Ref.~\onlinecite{DelaneyRinkeAuger}.
Our present results for the direct Auger process are in qualitative agreement with the work of
Bertazzi \emph{et al.}\cite{bertazzi:231118} who also observed a peak of the e--e--h Auger coefficient, but at a different energy ($E_{\text{gap}}=2.9$ eV).
We attribute the quantitative differences between our work and that of
Bertazzi \emph{et al.}\cite{bertazzi:231118} to the different band structures
(fully first-principles in pure GaN versus fitted pseudopotentials for virtual-crystal InGaN) and dielectric-function models used in the respective
calculations.
Finally, in the 1.0--1.5 eV energy range the e--e--h coefficient increases exponentially for decreasing band-gap values,
as expected for direct Auger recombination\cite{BulashevichKarpov},
because only direct intraband Auger transitions are possible for this band-gap range.

\begin{figure}
\includegraphics[width=\columnwidth]{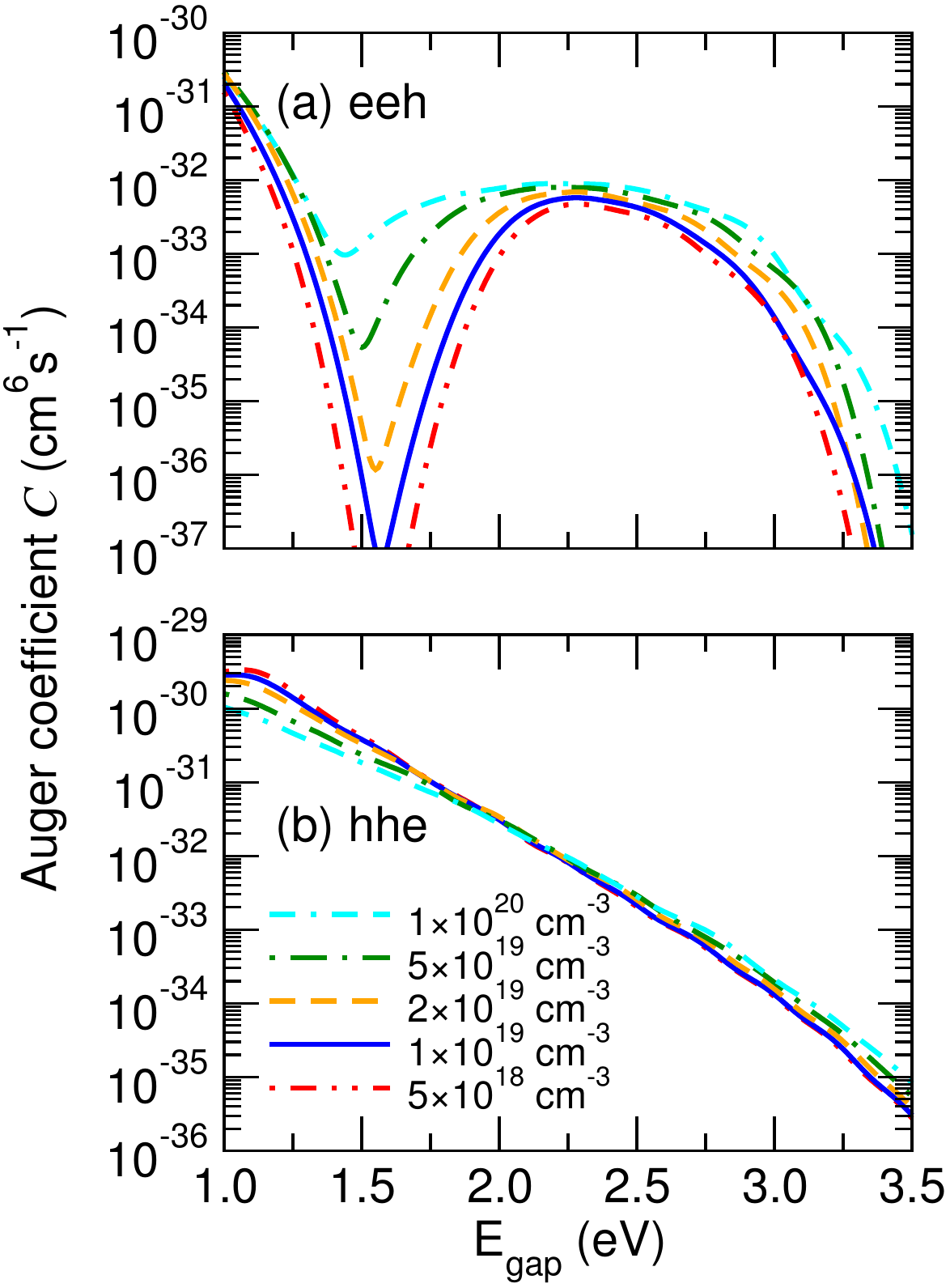}
\caption{\label{fig:gan_auger_vs_ndensity}
(Color online) Auger coefficients of GaN at 300 K due to (a) electron-electron-hole (e-e-h) and (b) hole-hole-electron (h-h-e) Auger
recombination processes plotted as a function of the scissors-shift-adjusted band gap for various free-carrier densities.}
\end{figure}

We also examined the dependence of the direct Auger coefficients on the free-carrier density.
Figures~\ref{fig:gan_auger_vs_ndensity}(a) and \ref{fig:gan_auger_vs_ndensity}(b) show the dependence of the e--e--h and h--h--e direct Auger processes
on the free-carrier density.
The dependence of the Auger coefficient on the carrier density is not monotonic as a function of the adjusted band gap.
For small adjusted band gaps (less than 2.0 eV)
the h--h--e  Auger coefficient is reduced for increasing carrier density.
This is due to phase-space filling effects\cite{Hader2005}.
Free carriers at low densities are nondegenerate and this results in a cubic carrier-dependence of the Auger rate.
For degenerate carrier densities, however, the Boltzmann approximation breaks down and carriers need to be described
by the Fermi-Dirac distribution, which reduces the cubic density dependence of the Auger rate.
In other words, the Auger coefficient becomes density-dependent at high carrier densities.
The Auger coefficient increases as a function of carrier density for larger vales of the adjusted band gap for h--h--e and for all adjusted band-gap values for e--e--h.
This is because direct Auger recombination for large values of the adjusted band gap is inhibited
by the limited density of states available for momentum-conserving Auger transitions.
However, as the carrier density increases the free carriers occupy a more extended region of the Brillouin zone
and more phase-space becomes available for Auger transitions.

\subsection{Alloy-scattering-assisted Auger recombination in In$_{0.25}$Ga$_{0.75}$N}
\label{sec:direct_ingan}

Alloying GaN with InN forms In$_x$Ga$_{1-x}$N alloys with band gaps desirable for optoelectronic applications.
In Sec.~\ref{sec:direct_gan} we modeled the resulting variation in Auger rates by changing the band gap.  However, alloying
also introduces an additional
carrier scattering mechanism due to the breaking of the translational periodicity by the substitution of Ga atoms with In.
As a result, the momentum-conservation requirement of Eq.~(\ref{eq:auger_me}) imposed by the crystal periodicity is relaxed for the alloy.
Consequently, Auger transitions that are forbidden in the perfect crystal because of the translational symmetry are allowed in the
alloy as a result of the substitutional disorder. The corresponding matrix elements acquire nonzero values.
By performing calculations for an InGaN alloy supercell,
which explicitly contains the effect of alloy disorder,
we show that alloying enables additional
Auger recombination processes that enhance the Auger coefficient.

\begin{figure}
\includegraphics[width=\columnwidth]{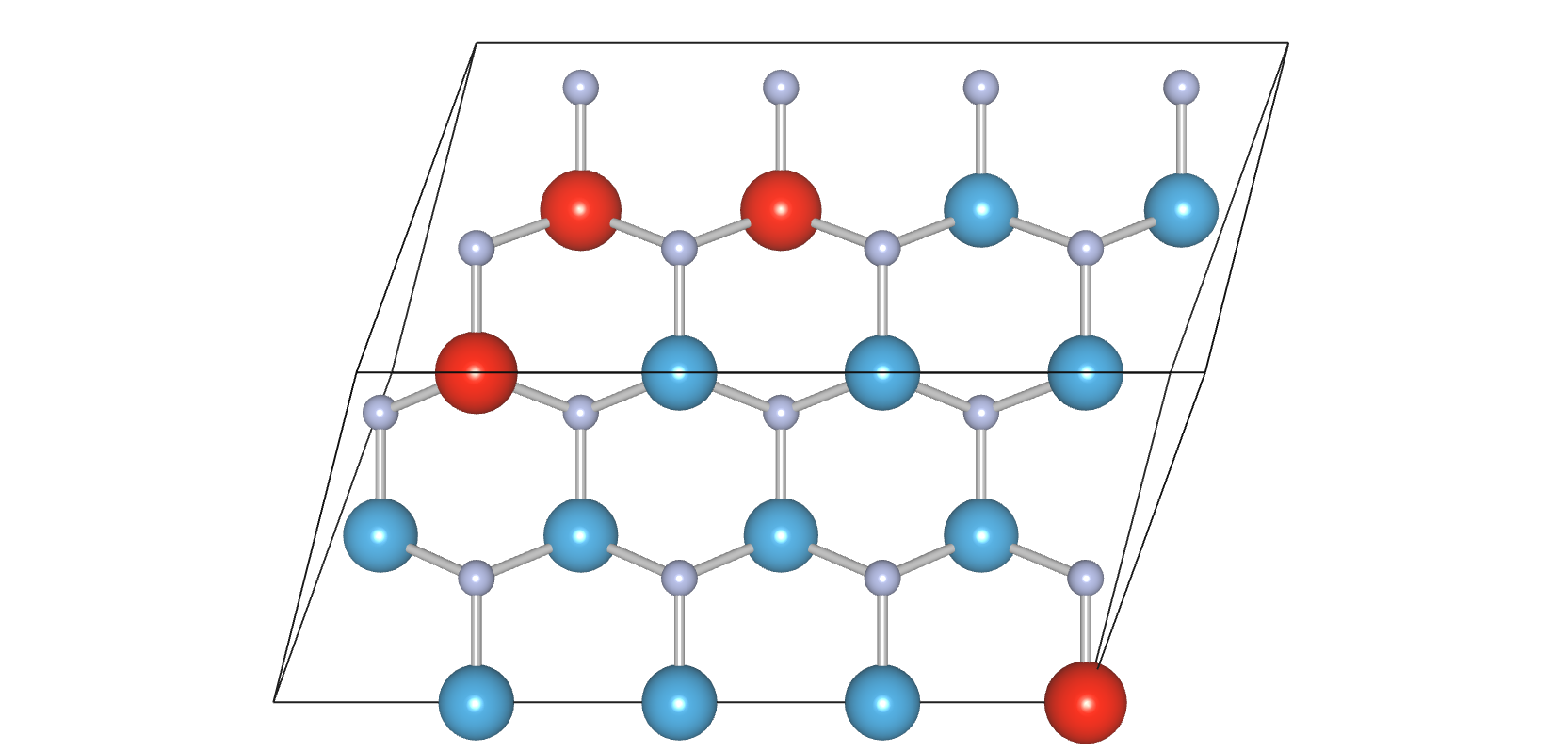}
\caption{
\label{fig:ingan25_sqs_structure}
(Color online) Structure of the quasirandom alloy cell used for the calculation of the Auger coefficient
in In$_{0.25}$Ga$_{0.75}$N (coordinates from Ref. \onlinecite{PhysRevB.74.024204}).
}
\end{figure}

Previous calculations of Auger recombination
in alloys used the virtual-crystal approximation to model the effect of alloying. However, the virtual crystal is a perfectly
periodic arrangement of atoms that misses important aspects of the actual alloy structure, such as substitutional disorder and atomic relaxations.
The importance of these two factors has been highlighted, for instance, in the case of TiO$_{2(a-x)}$S$_{2x}$, where they break the symmetry of optical matrix elements and enhance interband optical absorption,\cite{jp3106937} and also in the case of free-carrier absorption in In$_{x}$Ga$_{1-x}$N.\cite{PhysRevB.81.241201,APEX.3.082101}
In order to capture these effects in our simulations, we have to explicitly study Auger recombination in an alloy supercell that includes both kinds
of alloy atoms. A computationally feasible way to do this is with the 32-atoms special quasirandom alloy structure\cite{PhysRevLett.65.353} for the 25\% alloy
composition (Fig.~\ref{fig:ingan25_sqs_structure}) \cite{PhysRevB.74.024204}. This particular supercell has the
special property that it is the optimal 32-atom supercell that reproduces the short-range correlation function of the random alloy.

\begin{figure}
\includegraphics[width=\columnwidth]{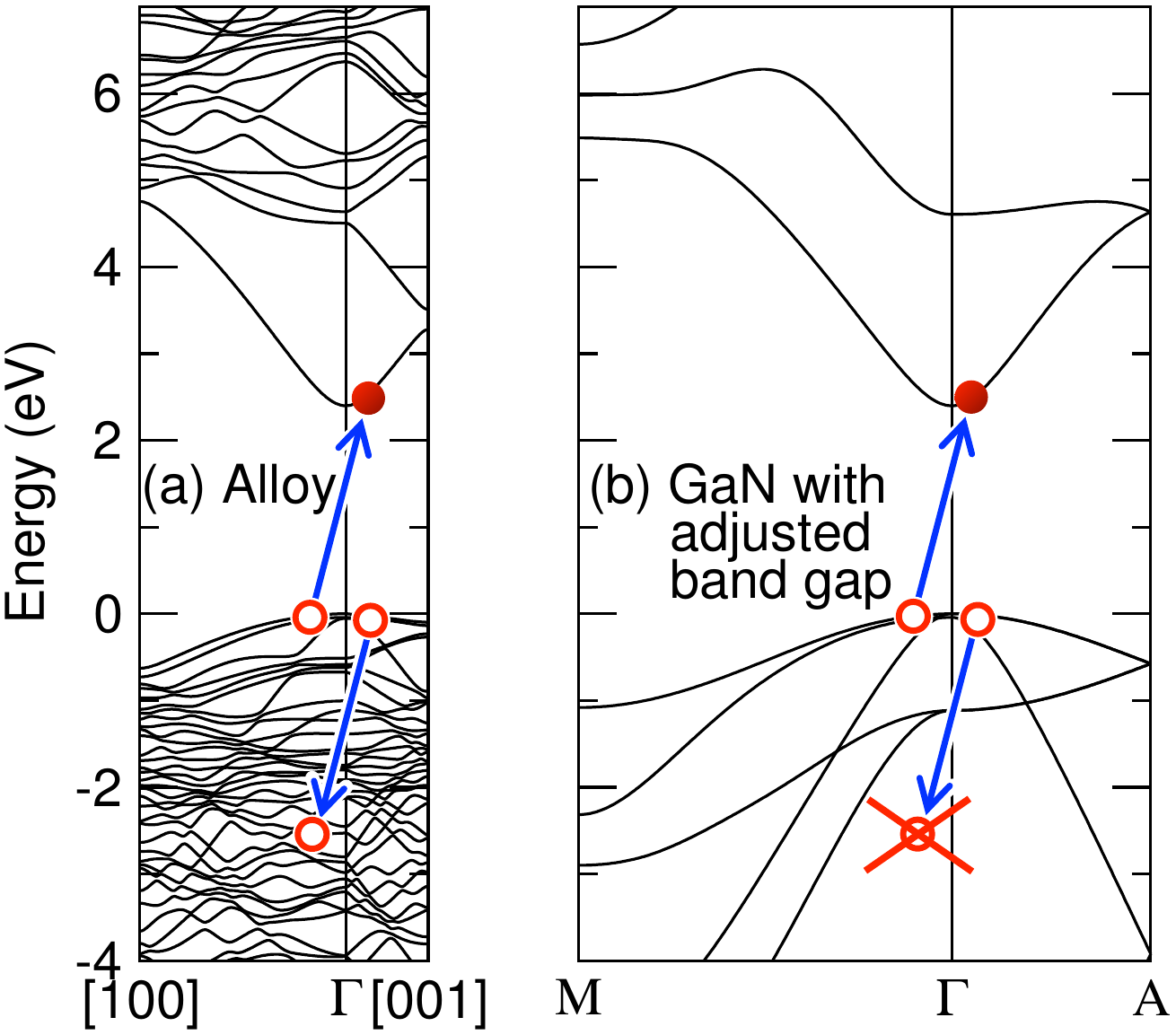}
\caption{
\label{fig:ingan_band_structure}
(Color online) Scissors-shift-adjusted band structure of In$_{0.25}$Ga$_{0.75}$N calculated using (a) the quasirandom alloy structure and
(b) bulk GaN with an adjusted band gap to match that of the alloy.
Alloying enables additional
Auger transitions, such as the illustrated h--h--e process,
that are not possible without the zone folding and symmetry breaking introduced by alloying.
}
\end{figure}

The scattering of free carriers by the alloy disorder is found to be important for Auger recombination and
significantly increases the Auger coefficients compared to pure GaN.
Figure~\ref{fig:ingan_band_structure}(a) shows the scissors-shift-adjusted band structure of In$_{0.25}$Ga$_{0.75}$N calculated for the 32-atoms alloy supercell
shown in Fig.~\ref{fig:ingan25_sqs_structure}, while Figure~\ref{fig:ingan_band_structure}(b) shows the band structure of GaN with a band gap adjusted to the
gap of In$_{0.25}$Ga$_{0.75}$N ($E_{\text{gap}}=2.4$ eV). The two figures show that the disorder introduced by alloying breaks the symmetry and
folds the band structure, which in turn enables additional Auger transitions and enhances the Auger coefficient.
These transitions
cannot be modeled
within the virtual-crystal approximation because the latter cannot describe the folded bands and the associated symmetry breaking. Our
first-principles calculations take into account the short-range alloy disorder and incorporate the effect of alloy scattering on
the Auger coefficients.

\begin{figure}
\includegraphics[width=\columnwidth]{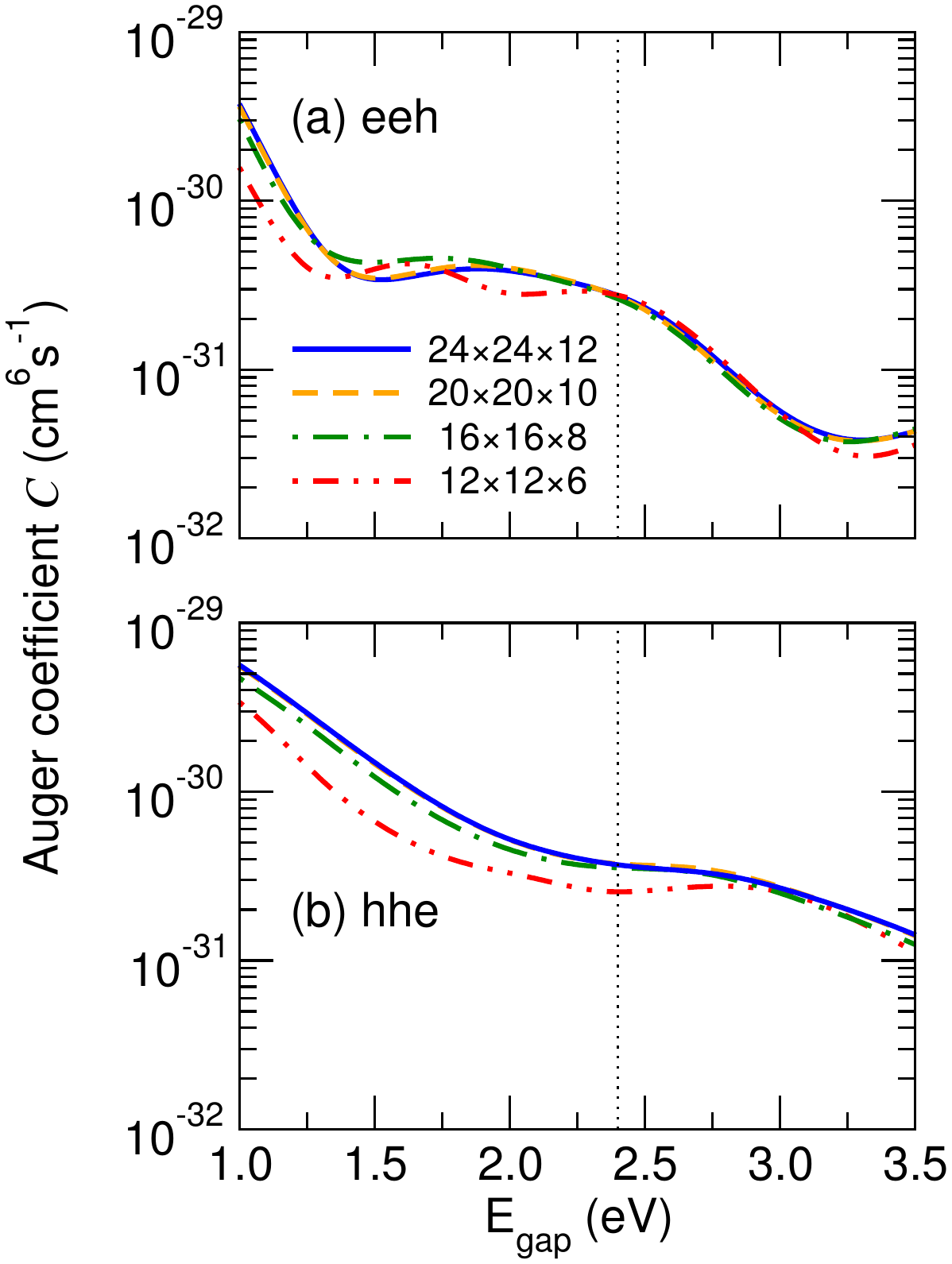}
\caption{\label{fig:ingan25_auger_vs_nk}
(Color online) Auger coefficients of In$_{0.25}$Ga$_{0.75}$N for a carrier density of 10$^{19}$ cm$^{-3}$ and a temperature of 300 K
due to (a) electron-electron-hole (e-e-h) and (b) hole-hole-electron (h-h-e) Auger
processes, plotted as a function of the band gap for various Brillouin-zone sampling densities.
The band gap is treated as an adjustable variable in order to model a varying alloy composition.
The variation of the alloy scattering potential with alloy composition on these Auger coefficients has not been included in this Figure,
but it is examined in Subsection~\ref{sec:total_auger}.}
\end{figure}

Once the additional scattering processes are taken into account, the dependence of the Auger rates on the band gap can again be examined by plotting results as a function of a scissors-shift-adjusted band gap, keeping the matrix elements fixed to those of the In$_{0.25}$Ga$_{0.75}$N alloy.
Figures~\ref{fig:ingan25_auger_vs_nk}(a) and (b) illustrate the dependence of the e--e--h and h--h--e alloy-scattering-assisted Auger coefficients
on the scissors-shift-adjusted band gap, as well as the convergence with respect to the Brillouin-zone-sampling
grid density.
The broadening of the delta function in Eq.~(\ref{eq:auger_rate}) has been set to 0.3 eV,
the free-carrier densities are $n=p=10^{19}$ cm$^{-3}$, and the temperature is 300 K.
The coefficients are on the order of 3$\times$10$^{-31}$ cm$^6$s$^{-1}$ for the adjusted gap value that
matches the 25\%-alloy band gap ($E_{\text{gap}}=2.4$ eV),
much larger than the corresponding values for GaN.

\begin{figure}
\includegraphics[width=\columnwidth]{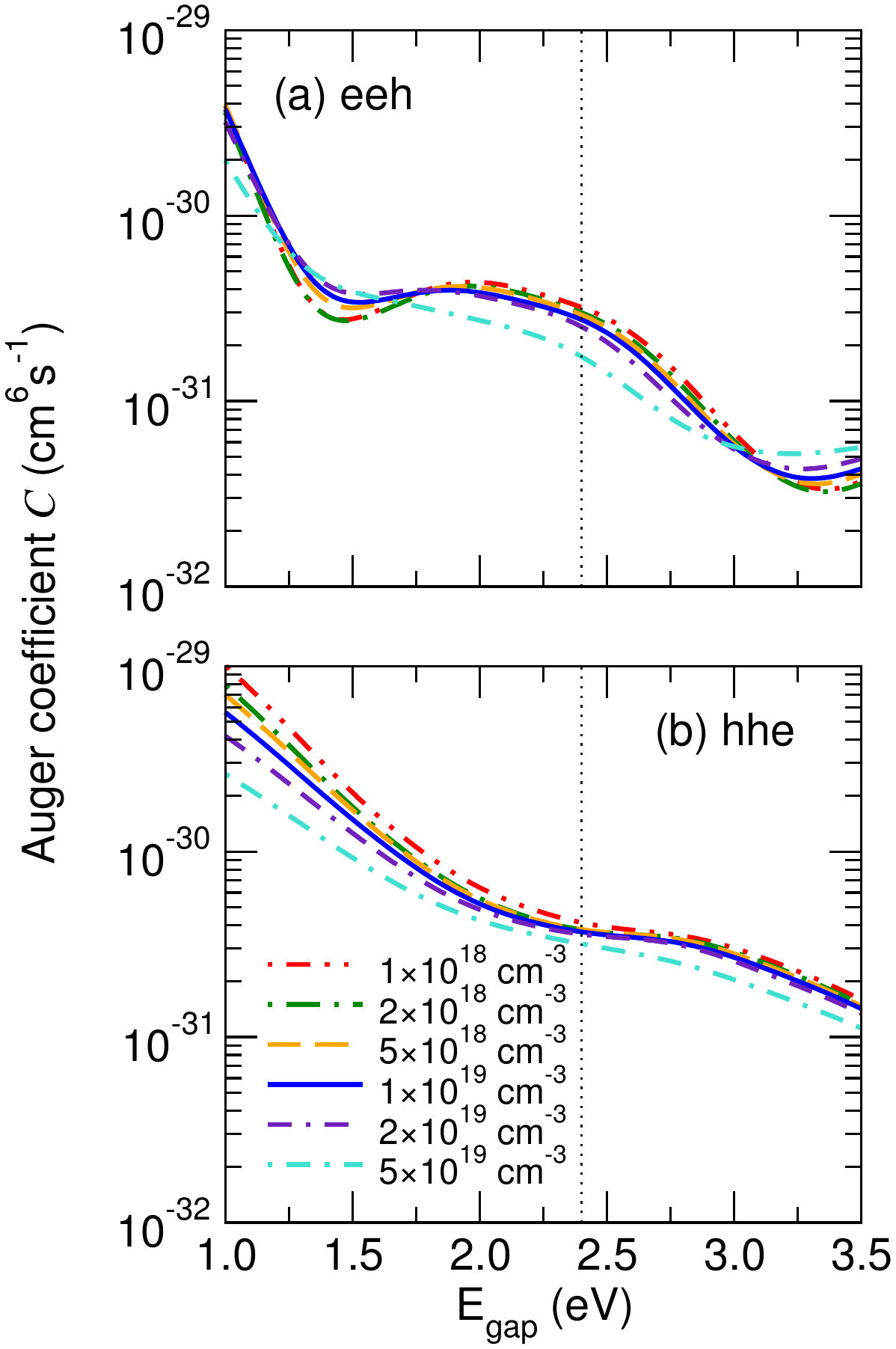}
\caption{\label{fig:ingan25_auger_vs_ndensity}
(Color online) Alloy-assisted Auger coefficients of In$_{0.25}$Ga$_{0.75}$N at 300 K due to (a) electron-electron-hole (e-e-h) and (b) hole-hole-electron (h-h-e) Auger
recombination processes, plotted as a function of the scissors-shift-adjusted band gap for various free-carrier densities.
The dotted line corresponds to the band gap of In$_{0.25}$Ga$_{0.75}$N (2.4 eV)
as predicted from hybrid functional calculations~\cite{Moses2010}. }
\end{figure}

The dependence of the alloy-assisted Auger coefficients on the free-carrier
density is shown in Figs.~\ref{fig:ingan25_auger_vs_ndensity}(a) and (b). Both the e--e--h and the h--h--e coefficients decrease
for increasing densities for every value of the adjusted band gap. This implies that for the alloy-assisted Auger case,
the effects of phase-space filling are more important than the enabling of additional Auger transitions for higher carrier densities.
This is expected, since bands farther from the zone center have already been made accessible for direct Auger transitions by the zone folding due to alloying.

\begin{figure}
\includegraphics[width=\columnwidth]{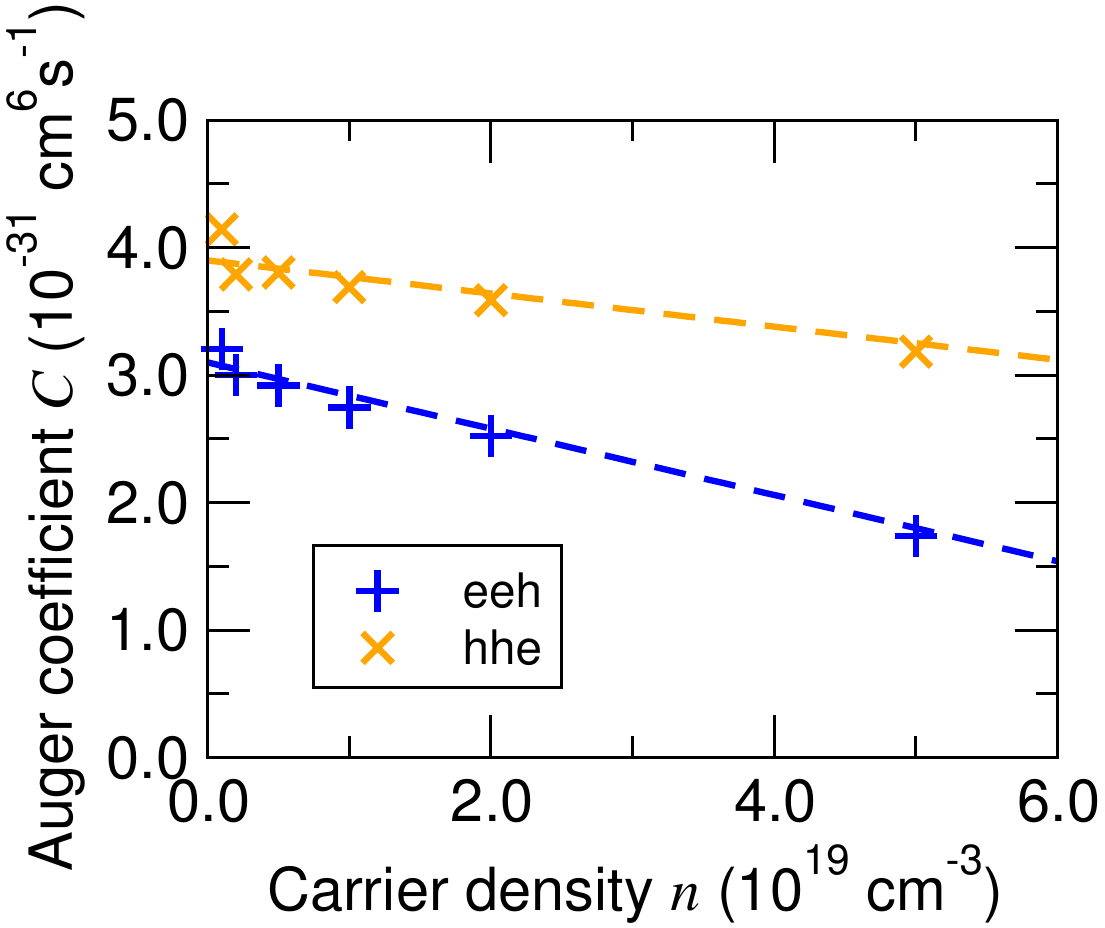}
\caption{\label{fig:auger_vs_n}
(Color online) Carrier-density dependence of the alloy-scattering-assisted Auger coefficients of In$_{0.25}$Ga$_{0.75}$N at 300 K.
The coefficients decrease for increasing carrier densities due to the transition from nondegenerate to degenerate carrier statistics (phase-space filling).}
\end{figure}

The density dependence of the alloy-scattering-assisted Auger coefficients of In$_{0.25}$Ga$_{0.75}$N at 300 K for an adjusted band-gap value of 2.4 eV
is plotted as a function of carrier concentration in Fig.~\ref{fig:auger_vs_n}.
The values of the Auger coefficients decrease linearly for increasing carrier densities
in the $2\times 10^{18}-5\times10^{19}$ cm$^{-3}$ range, which is the typical range of free-carrier concentrations under LED operating conditions.
The decrease of the Auger coefficients with increasing density is due to the transition from nondegenerate to degenerate carrier statistics (phase-space filling) in this carrier-density regime and to the increasing screening of the Coulomb interaction matrix elements by free carriers.
We fit the first-principles data in the $2\times 10^{18}-5\times10^{19}$ cm$^{-3}$ carrier-concentration range with a linear equation of the form
\begin{equation}
C(n)/10^{-31} \text{cm}^6\text{s}^{-1}= a + b n/10^{19}\text{cm}^{-3},
\end{equation}
where $a$ and $b$ are dimensionless fitting parameters with values $a=3.1$, $b=-0.26$ for e--e--h and $a=3.9$, $b=-0.13$ for h--h--e.

We note that the calculated density dependence of the alloy-scattering-assisted Auger coefficients differs from the expression
$C(n)=C_0/(1+n/n_0)$
which has been used in the literature to describe the effect of phase-space filling\cite{Hader2005,David2010a}.
Our present results show that the rate of decrease of the Auger coefficient with density in the $2\times 10^{18}-5\times10^{19}$ cm$^{-3}$ range
(which includes the peak of the internal quantum efficiency of LEDs\cite{David2010b})
is smaller than the rate of decrease of the radiative coefficient ($B$)\cite{Kioupakis2013}.
The more rapid decrease of the $B$ coefficient in comparison to the $C$ coefficient with increasing free-carrier density
may explain the asymmetry of the internal quantum efficiency curve
reported in the literature,\cite{Dai2011} without invoking an additional fourth-order or higher-power carrier-loss mechanism.


\subsection{Phonon-assisted Auger recombination in GaN}
\label{sec:phonon_gan}

\subsubsection{Phonon-assisted Auger coefficients}

\begin{figure}
\includegraphics[width=\columnwidth]{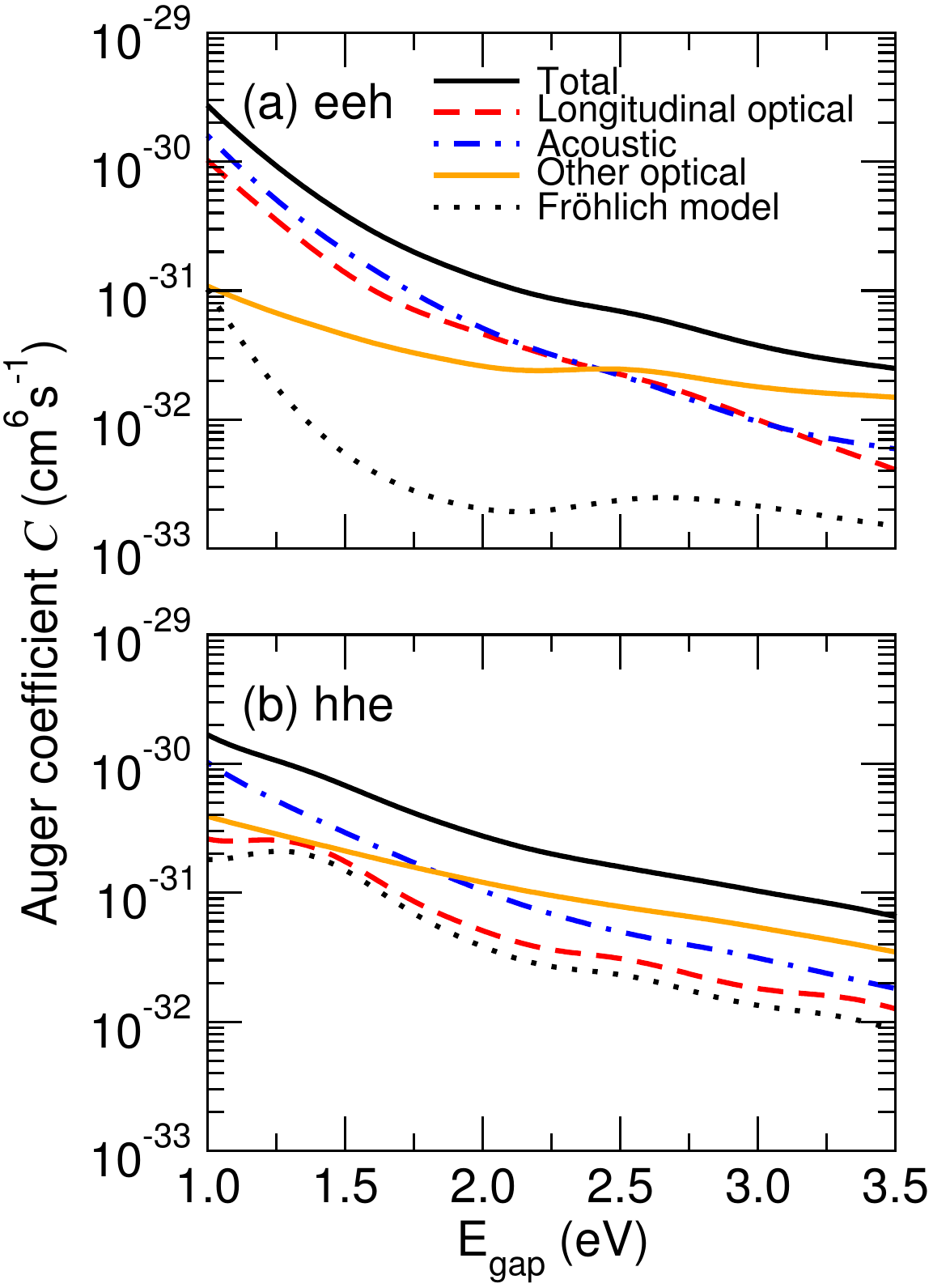}
\caption{\label{fig:auger_phonon_bymode}
(Color online) Contributions by the various phonon modes [longitudinal optical (LO), acoustic, and other optical modes] to the total phonon-assisted Auger coefficient of GaN for the (a) e--e--h and (b) h--h--e processes at 300 K. The phonon-assisted Auger coefficient using the Fr\"ohlich model for the electron-phonon coupling is also shown for comparison.}
\end{figure}

Figures~\ref{fig:auger_phonon_bymode}(a) and \ref{fig:auger_phonon_bymode}(b) display our calculated results of the phonon-assisted Auger coefficients at 300 K
for the e--e--h and h--h--e  processes, respectively.
To model the various alloy compositions we again used the calculated parameters for GaN and adjusted the band gap with a rigid scissors shift.
By varying the band gap, we can model transitions to different sets of final states and this enables us to
model phonon-assisted Auger recombination for a range of In$_x$Ga$_{1-x}$N compositions.
The broadening of the $\delta$ function in Eqs.~(\ref{eq:indirect_auger_rate_Gamma}) and~(\ref{eq:indirect_auger_rate_hhe_Gamma}),
as well as the imaginary component of the energy denominator in Eq.~(\ref{eq:gme_Gamma}) were set to 0.3 eV.
The phonon-assisted Auger coefficients are much larger than the values of the direct Auger coefficient in GaN.
The coefficients increase for decreasing band-gap values, indicating that Auger recombination is enhanced for higher-In-composition alloys.

\subsubsection{Phonon mode and wave vector contributions to phonon-assisted Auger coefficients}

One interesting question to examine is which phonons contribute
to the Auger process and how well their contributions to the Auger processes
are captured with more approximate theoretical treatments, as opposed to the first-principles density functional theory treatment used in the present work.
The contributions from each type of vibrational mode to the Auger coefficient at 300 K
are also plotted in Figs.~\ref{fig:auger_phonon_bymode}(a) and \ref{fig:auger_phonon_bymode}(b) for the e--e--h and h--h--e processes, respectively.
For both kinds of Auger processes, and for band-gap values in the 2.4--3.5 eV range, the dominant electron-phonon processes
are due to optical-phonon deformation-potential scattering,
followed by scattering by the acoustic phonon modes and the polar optical phonons.
Figures~\ref{fig:auger_phonon_bymode}(a) and \ref{fig:auger_phonon_bymode}(b) also include a comparison of the first-principles results
to the phonon-assisted Auger coefficients calculated using the Fr\"ohlich expression for the electron-phonon interaction potential,
which models the scattering of carriers by polar interactions with the longitudinal-optical phonons\cite{Ridley}
and has been shown to agree with the first-principles results at long wavelengths\cite{PhysRevB.81.241201}:
\begin{equation*}
\Delta V_{\text{el-ph}} = -\frac{i}{q}\left[ \frac{2\pi e^2 \hbar \omega_{\text{LO}} }{V} \left( \frac{1}{\varepsilon_{\infty}}-\frac{1}{\varepsilon_0} \right) \right]^{1/2}.
\end{equation*}
For the model parameters, we used
$\varepsilon_{0}=(\varepsilon_{0\parallel}\varepsilon_{0\perp}^2)^{1/3}=9.79$\cite{BarkerIlegems1973},
$\varepsilon_{\infty}=(\varepsilon_{\infty \parallel}\varepsilon_{\infty \perp}^2)^{1/3}=5.5$\cite{BarkerIlegems1973},
and $\hbar\omega_{\text{LO}}=92$ meV\cite{giehler:733}.
For band-gap values in the 2.4--3.0 eV energy range, the Auger coefficients calculated with the Fr\"ohlich model are smaller than the first-principles ones by approximately one order of magnitude. For the h--h--e process the Fr\"ohlich model agrees well with the first-principles data for the LO-phonon contribution to the scattering, but for the e--e--h case the Fr\"ohlich model cannot account even for that contribution. As we will show later, the reason for this discrepancy is because the phonons involved in the scattering process for band-gap values in the 2.4--3.0 eV range are short-ranged ones (i.e., they involve large momentum transfer) and cannot be captured by the Fr\"ohlich model, which is derived for the long-wavelength limit.

\subsubsection{Limitations of approximate band-structure and phonon models}

\begin{figure}
\includegraphics[width=\columnwidth]{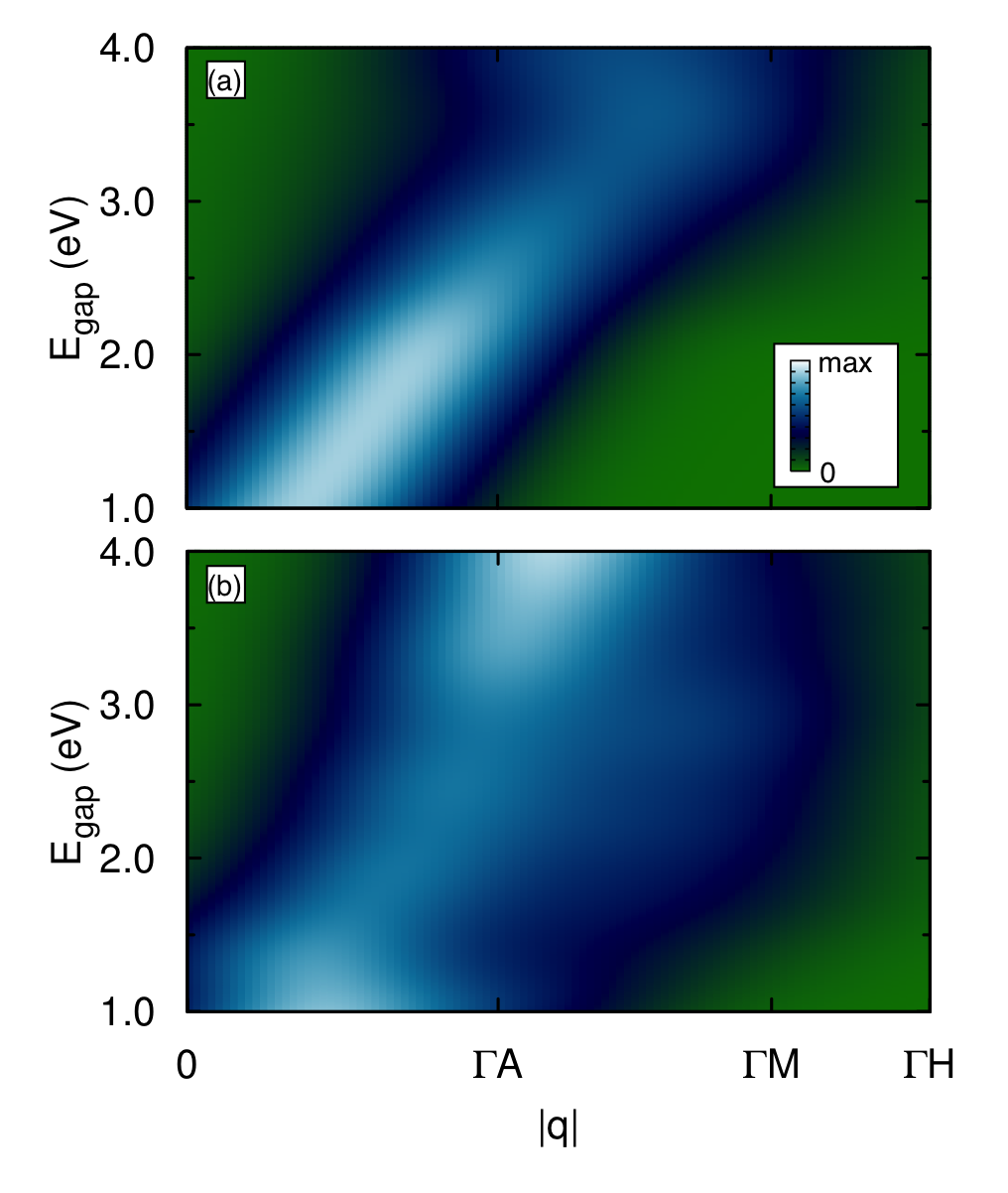}
\caption{\label{fig:phonon_byq}
(Color online) Relative magnitude of phonon-assisted Auger coefficients of GaN at 300 K due to
(a) electron-electron-hole (e-e-h) and
(b) hole-hole-electron (h-h-e) Auger recombination processes,
plotted as a function of the scissors-shift-adjusted band gap and analyzed in terms of the norm of the q-points that contribute to the recombination process.
The norms of the vectors from the $\Gamma$ $(0,0,0)$ point to the A $(0,0,\frac{1}{2})$, M $(\frac{1}{2},0,0)$, and H $(\frac{2}{3},\frac{1}{3},\frac{1}{2})$ points
are denoted along the horizontal axis. The coordinates of the points are given in relative coordinates with respect to the reciprocal lattice vectors.}
\end{figure}

The limitations of approximate phonon models are further illustrated by an analysis of the wave vectors
of the phonons that provide the dominant contribution to phonon-assisted Auger recombination.
Figure~\ref{fig:phonon_byq}
shows the contributions to the phonon-assisted Auger coefficients from phonons with different momenta $\bm{q}$
as a function of the adjusted band gap (to model the effect of alloying, but using parameters for GaN) and the norm of the phonon momentum $|\bm{q}|$.
The plots indicate that for band gap values in the 2.4--3.0 eV range,
the dominant contributions to the Auger coefficient originate from scattering by phonons
with momenta comparable to the Brillouin-zone dimensions [i.e., comparable
to the distance from the $\Gamma$ to the A point at the boundary of the first Brillouin zone of wurtzite (Fig.~\ref{fig:ingan_band_structure})].
This indicates that the phonon-scattering events involved in the indirect Auger process
in nitride devices with wavelengths from violet to green are short ranged.
This is dictated by the band structure and by the fact that the higher conduction- (valence-)band states, which
accommodate the electrons (holes) excited
by the e--e--h (h--h--e) processes, are located at wave vectors
comparable to the size of the first Brillouin zone.

Our analysis indicates that since phonon-assisted Auger recombination is a short-range effect,
it will not be significantly affected by
quantum confinement.
The values of the phonon-assisted Auger coefficients
for quantum wells with a width of 2.5--3 nm (i.e., 5--6 times the lattice constant along the \emph{c} axis), which is typical for commercial nitride LEDs, are expected to be approximately equal to their bulk values.
Our results also show that the values of the phonon-assisted Auger coefficients
cannot significantly be suppressed by engineering approaches such as nanopattering,
because these methods do not modify materials properties on the scale of the unit cell.

These results also validate the approximation of using $\Gamma$-point-only wave functions for the
initial electron and hole states.
It is known from $\bm{k}\cdot \bm{p}$ theory that the difference between the wave functions at $\Gamma$ and at a $\bm{k}$-point in the vicinity of $\Gamma$ is
proportional to $|\bm{k}|$\cite{YuCardona}. Since matrix elements are proportional to the square of the wave function,
the error we make by approximating the initial-state wave functions of the carriers with those at $\Gamma$ is on the order of the square of the ratio of the
free-carrier Fermi wave vector versus the dominant-phonon wave vectors, i.e., the Brillouin-zone dimensions.
For $n=10^{19}$ cm$^{-3}$, the free-carrier Fermi wave vector at $T=0$ K is on
the order of 0.03 $a_B^{-1}$, while the Brillouin-zone dimension (distance from $\Gamma$ to $A$ point) is 0.64 $a_B^{-1}$.
Therefore, the error introduced by this approximation is on the order of 0.2\%,
which is sufficient for our purposes.

Another conclusion that can be drawn from the results of Figs.~\ref{fig:auger_phonon_bymode} and \ref{fig:phonon_byq} is that
the calculations using $\bm{k}\cdot\bm{p}$ theory for the band structure and the Fr\"ohlich model for the
electron-phonon coupling employed in Ref.~\onlinecite{Pasenow2009} cannot properly capture
Auger recombination in wide-band-gap nitride materials.
This is because $\bm{k}\cdot\bm{p}$ theory is inaccurate for the bands and wave functions of states in the energy range
at 2.4--3.0 eV away from the band edges, which is relevant for Auger recombination in the nitrides,
and therefore cannot accurately include the contribution of these states to the overall recombination rate.
Moreover, Figs.~\ref{fig:auger_phonon_bymode}(a) and (b) illustrate that the dominant contributions to the phonon-assisted Auger recombination rate
arises from deformation-potential optical phonon scattering and from acoustic phonons, while the LO phonons contribute only
a small fraction to the overall rate.
In addition, the results in Figs.~\ref{fig:auger_phonon_bymode}(a) and (b) show that the Fr\"ohlich model cannot accurately capture the contribution
of the LO phonons to the Auger rate and that a microscopic description of electron--LO-phonon coupling is necessary.
The proper description of the energies and wave functions of states away from the band edges
as well as the incorporation of all phonon modes in electron-phonon scattering provided by the first-principles calculations of this work
is essential to obtain accurate results
for phonon-assisted Auger recombination in the nitrides.

\subsubsection{Appropriateness of second-order perturbation theory}

Our results also show that another previous study of phonon-assisted Auger recombination in the nitrides\cite{bertazzi:011111},
which is based on the formalism of Bardyzewski and Yevick\cite{bardyszewski:2713}, does not incorporate
the dominant phonon-assisted processes and significantly underestimates the magnitudes of the phonon-assisted Auger coefficients in the nitrides.
The method used in Ref.~\onlinecite{bertazzi:011111} incorporates the effect of electron-phonon coupling in the broadening of the carrier spectral
functions, but it does not consider the additional momentum provided to the carriers by phonon emission or absorption
[Eq.~(3) in Ref.~\onlinecite{bertazzi:011111}]. In other words, the indirect Auger processes facilitated by the additional momentum provided by the phonons
that excite carriers to higher electronic states near the edges of the Brillouin zone, which we found to dominate phonon-assisted Auger recombination,
were not considered in Ref.~\onlinecite{bertazzi:011111}. We therefore attribute the difference between our results and the ones reported in Ref.~\onlinecite{bertazzi:011111}
to the omission of the dominant phonon terms in the latter work.

\begin{figure}
\includegraphics[width=\columnwidth]{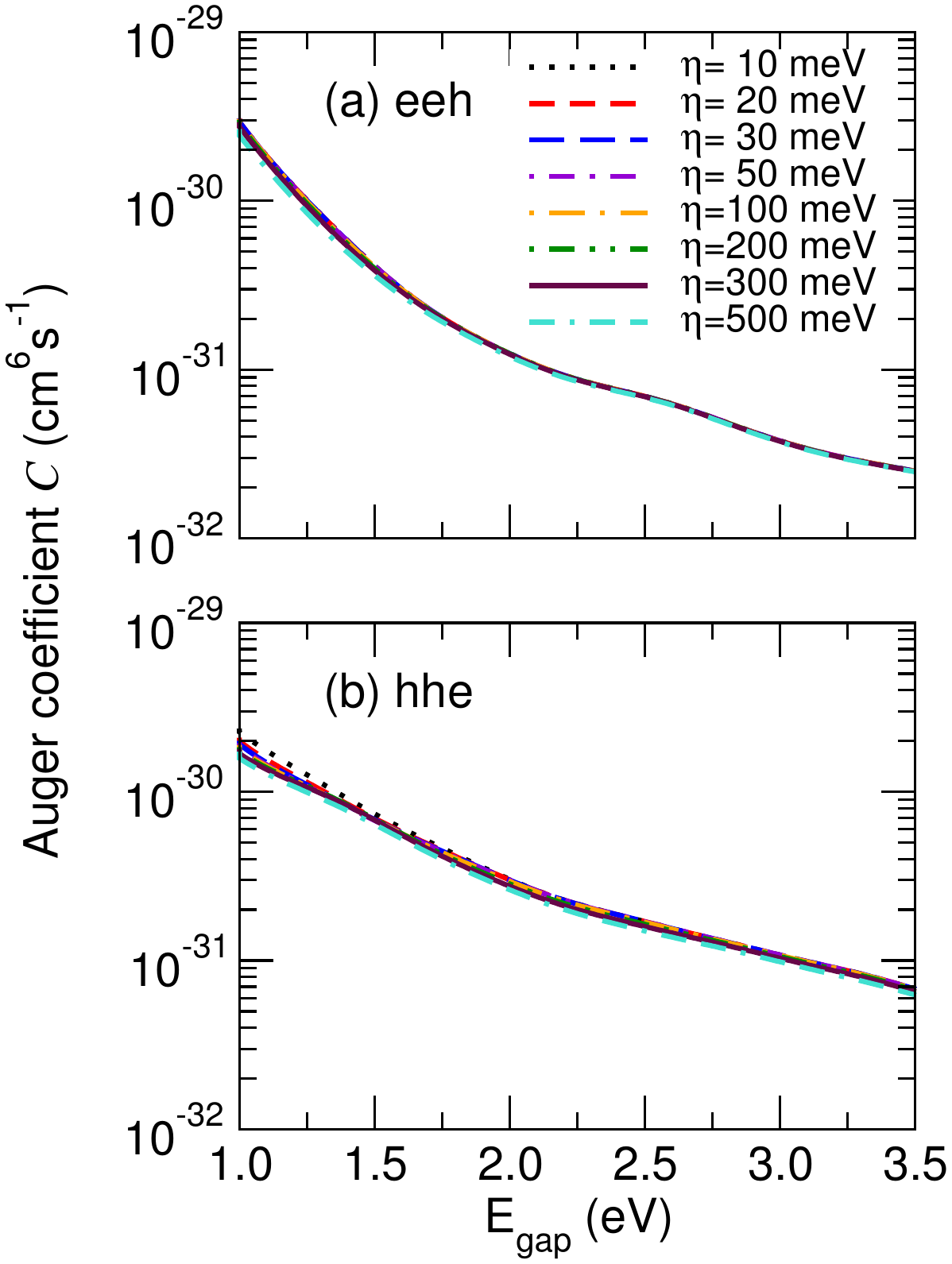}
\caption{\label{fig:auger_lifetime}
(Color online) Phonon-assisted Auger coefficient as a function of the scissors-shift-adjusted band gap of GaN for various
values of the energy broadening parameter $\eta$ in the energy denominator of Eq.~(\ref{eq:gme_Gamma}).
}
\end{figure}

It was also argued in Ref.~\onlinecite{bertazzi:011111} that a treatment of phonon-assisted Auger recombination with second-order perturbation theory as performed
in the present work may be inadequate because of the need to use the imaginary broadening parameter ($\eta$) in the denominator of the matrix elements in Eq.~(\ref{eq:gme_Gamma}).
We have explicitly examined the dependence of the calculated phonon-assisted e--e--h and h--h--e Auger coefficients
on the value of the broadening parameter $\eta$ used in Eq.~(\ref{eq:gme_Gamma}).
Figure~\ref{fig:auger_lifetime} shows that the calculated coefficients
remain approximately constant as the broadening parameter varies over almost two orders of magnitude.
For adjusted band-gap values in the 2.4--3.5 eV range, the Auger coefficients change only by 2\% for the e--e--h and by 14\% for the h--h--e processes as the broadening parameter varies by a factor of 50.  The dependence on the broadening parameter is slightly greater for band-gap values in the 1.0--2.0 eV range, but still the e--e--h coefficient changes by at most 21\% and the h--h--e coefficient by 43\% at most.

The imaginary part of the energy denominator
may in general affect the transition rates calculated with second-order perturbation theory if, as pointed out in Ref.~\onlinecite{bertazzi:011111},
resonant direct transitions to the intermediate states are possible.
However, this is not the case for phonon-assisted Auger recombination in the nitrides. The intermediate states involved
are far from being resonant and the real part of the energy denominator of the matrix elements in Eq.~(\ref{eq:gme_Gamma}) dominates.
Therefore, the theoretical treatment of phonon-assisted Auger recombination in terms of perturbation theory yields results that do not significantly depend
on the chosen value of the imaginary broadening parameter.
We also note that for materials for which resonant intermediate states do matter,
the imaginary part of the energy denominator induced by electron-phonon coupling can
be calculated with first-principles methods (e.g., Ref.~\onlinecite{PhysRevLett.99.086804})
and incorporated in our formalism.

\subsubsection{Temperature dependence of phonon-asissted Auger coefficients}

\begin{figure}
\includegraphics[width=\columnwidth]{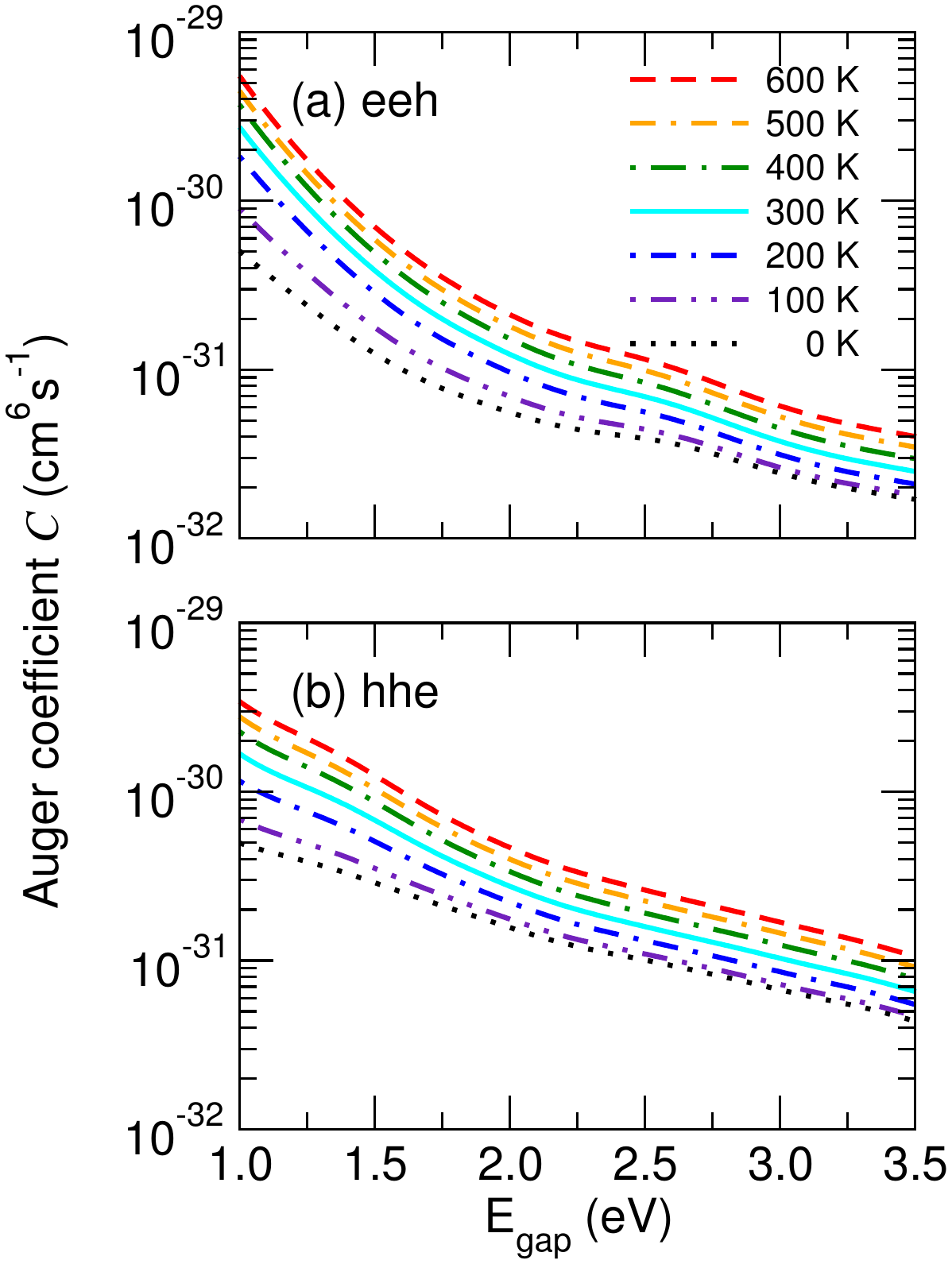}
\caption{\label{fig:auger_phonon_vs_T}
(Color online) Temperature dependence of the (a) e--e--h and (b) h--h--e phonon-assisted Auger coefficient of GaN as a function of the scissors-shift adjusted band gap.
}
\end{figure}

Figure~\ref{fig:auger_phonon_vs_T} shows the temperature dependence of the phonon-assisted Auger coefficients of GaN as a function of the scissors-shift-adjusted band gap.
The Auger coefficients increase at higher temperatures because the phonon occupation numbers in Eq.~(\ref{eq:indirect_auger_rate}) are a monotonically increasing
function of temperature. We note that phonon-assisted Auger recombination is possible even at absolute zero temperature because the phonon-emission-assisted
process does not require a finite thermal phonon population.

Our calculated temperature dependence of the Auger coefficient is in agreement with experimental measurements.
Galler \emph{et al.}\cite{galler:131111} measured the radiative and Auger coefficients for a single-quantum-well InGaN device.
After accounting for the reduced overlap due to polarization fields, they determined the values of the Auger coefficient as a function of temperature.
Their data indicates that the Auger coefficient in InGaN is a monotonically increasing function of temperature, in agreement with our findings for
phonon-assisted Auger recombination.
Moreover, the internal-quantum-efficiency data reported by Laubsch \emph{et al.}\cite{Laubsch2009} show that Auger recombination in InGaN
occurs even at cryogenic temperatures (4 K).
This is in contrast with the expected exponential temperature dependence of direct Auger recombination\cite{Ridley} but
in agreement with our finding that phonon-emission-assisted Auger processes are possible even at absolute zero temperature.

\subsubsection{Influence of strain on phonon-asissted Auger coefficients}

\begin{figure}
\includegraphics[width=\columnwidth]{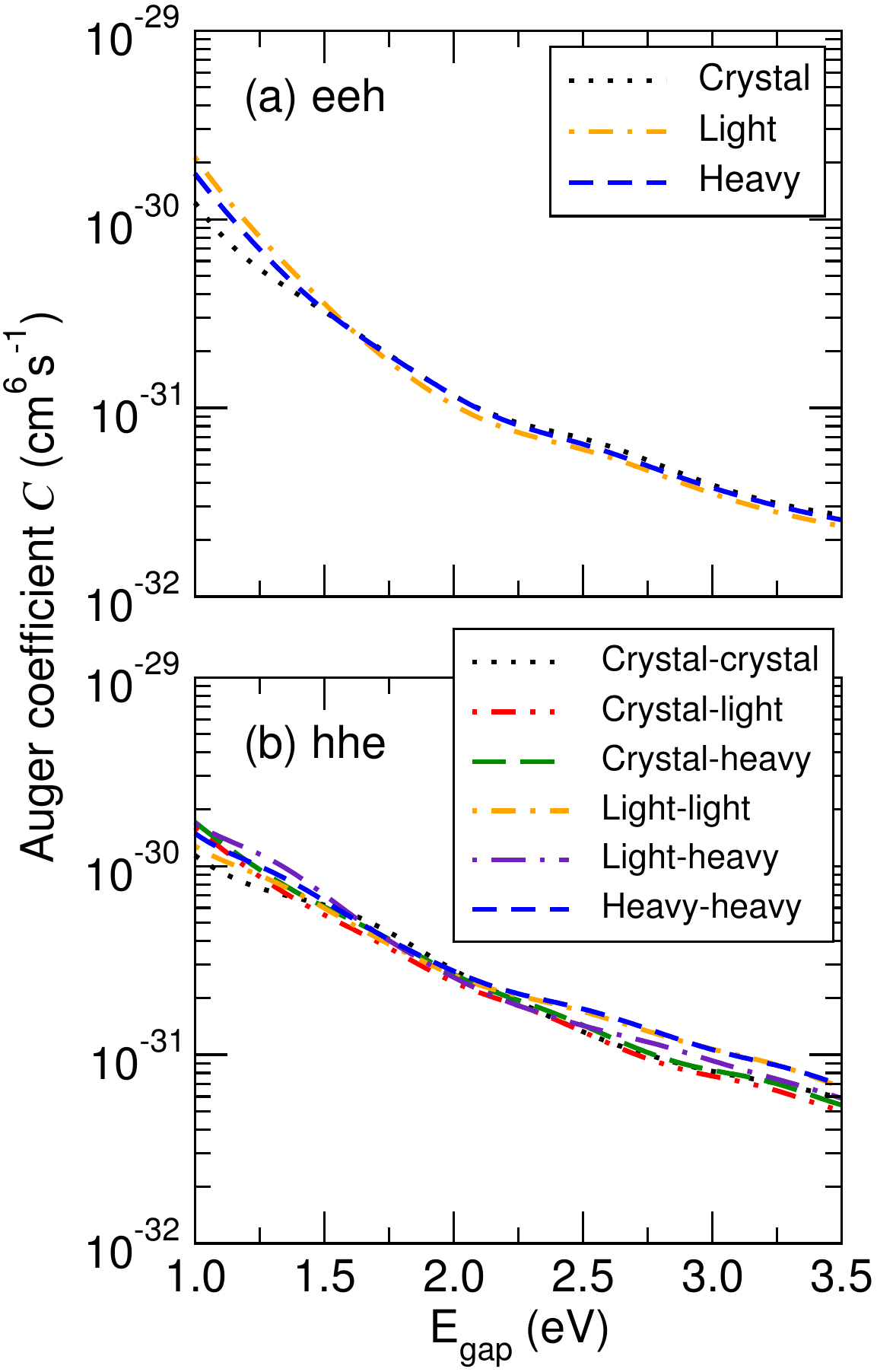}
\caption{\label{fig:auger_strain}
(Color online) Phonon-assisted Auger recombination in strained GaN as a function of the scissors-shift-adjusted band gap for the
(a) electron-electron-hole (eeh) and (b) hole-hole-electron (hhe) processes.
The various curves correspond to the various hole states (heavy hole, light hole, or crystal-field-split hole)
that participate in the Auger transitions.
}
\end{figure}

We also explored the influence of strain on the value of the phonon-assisted Auger coefficients, but we found that the effect is very small.
Strain modifies the relative energy of the valence bands near the maximum at $\Gamma$ \cite{yan:121111}
and influences the electron-hole recombination.
It is therefore conceivable that certain strain conditions could yield a favorable valence-band ordering that minimizes the Auger coefficients.
To test this hypothesis, we calculated the phonon-assisted Auger coefficients
in strained GaN.
We applied a 0.5\% tensile strain along one of the primitive vectors in the \emph{c}-plane of the wurtzite structure, which breaks the valence-band degeneracy and allows us
to distinguish transitions involving various hole states. The value of the Auger coefficients for various InGaN compositions was again calculated by applying a
rigid scissors shift to the band gap. The calculated results are shown in
Fig.~\ref{fig:auger_strain}. The e--e--h phonon-assisted Auger coefficient depends
weakly on the choice of hole state [Fig.~\ref{fig:auger_strain}(a)]. This is expected by symmetry
since electrons occupy Ga \emph{s} states, while the top of the valence band is made of N \emph{p} orbitals oriented along different directions.
On the other hand, we expect variations
for the h--h--e
coefficient depending on whether the two holes occupy the same or different valence bands.
Our results, however, show that this dependence is weak [Fig.~\ref{fig:auger_strain}(b)].
Therefore, the aplication of strain in order to reorder the valence bands
and alter the hole state occupations does not
significantly reduce the value of the phonon-assisted Auger recombination rate,
which is the primary Auger recombination mechanism in the nitrides.

\subsection{Charged-defect-assisted Auger recombination in GaN}
\label{sec:defect_gan}

\begin{figure}
\includegraphics[width=\columnwidth]{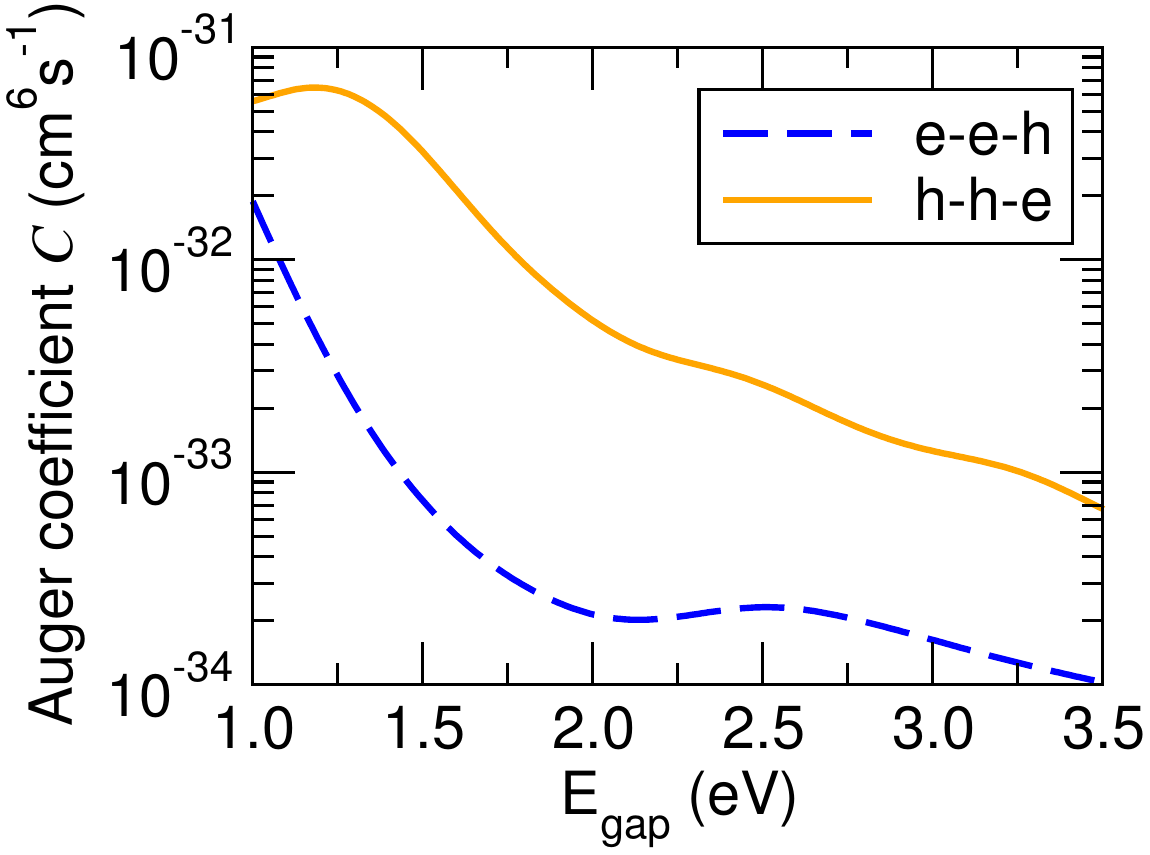}
\caption{\label{fig:auger_chargedefects}
(Color online) Charged-defect-assisted Auger recombination in GaN for a charged-defect concentration of 10$^{19}$ cm$^{-3}$.}
\end{figure}

Another scattering mechanism that can in principle contribute to
indirect Auger recombination is scattering due to charged defects.
The formalism for the calculation of the charged-defect-assisted Auger coefficient is similar to the phonon-assisted case, except the phonon frequencies in Eqs.~(\ref{eq:indirect_auger_rate}) are set to zero, and the electron-phonon scattering matrix elements ($g$) to the corresponding charged-defect-scattering ones, $g_{\text{defect}}=\langle \bm{k} | \Delta V_{\text{defect}} | \bm{k}+\bm{q} \rangle$, where $\Delta V_{\text{defect}}$ is the screened Coulomb potential of a defect of charge $Z$:
\[ \Delta V_{\text{defect}} = \frac{4\pi Z e^2}{\varepsilon(q)(q^2+\lambda^2)}. \]
The calculations were performed using band-structure and screening parameters for GaN and rigidly shifting the band gap to model the effect of alloying.
We investigated the scattering induced by negatively charged defects on Ga sites.  The results for the e--e--h and h--h--e processes for a free-carrier concentration of 10$^{19}$ cm$^{-3}$ are shown in Fig.~\ref{fig:auger_chargedefects}. We assumed a singly charged defect ($Z=1$) with a density of 10$^{19}$ cm$^{-3}$.   This density is actually much higher than observed defect densities, for instance for the Ga vacancies that are the most common point defects in $n$-type GaN\cite{VandeWalleNeugebauer_2004,Tuomisto201293}.  Figure~\ref{fig:auger_chargedefects} illustrates that charged-defect-assisted Auger recombination is weak for realistic densities of charged defects.
Even if we assume a concentration of 10$^{19}$ cm$^{-3}$ of triply charged defects ($Z=3$, the charge state expected for gallium vacancies in $n$-type GaN), the calculated charged-defect-assisted Auger coefficients increase only by a factor
of 9 and are still smaller by one order of magnitude compared to the phonon-assisted results.
In order to get an effect comparable to phonon-assisted Auger, the defect density would need to be on the order of 10$^{21}$ cm$^{-3}$ of singly charged defects or 10$^{20}$ cm$^{-3}$ of triply charged ones, which are unrealistically high and not encountered in actual devices. Therefore, charged-defect-assisted Auger recombination is an unlikely candidate to explain the efficiency droop in nitride light emitters. This is in agreement with experimental findings that the Auger coefficient does not change significantly for samples of varying quality\cite{Shen_et_al_2007}.
We note that our results only included the effect of long-range Coulomb scattering by charged impurities, and do not include effects such as central-cell corrections\cite{Ralph1975}, short-range scattering by neutral defects, or Auger transitions to localized defect states\cite{Hangleiter1987}. Further work that considers these effects is needed to fully evaluate the interplay of defects and Auger recombination in nitride materials.

\subsection{Cumulative Auger recombination coefficients}
\label{sec:total_auger}

\begin{figure}
\includegraphics[width=\columnwidth]{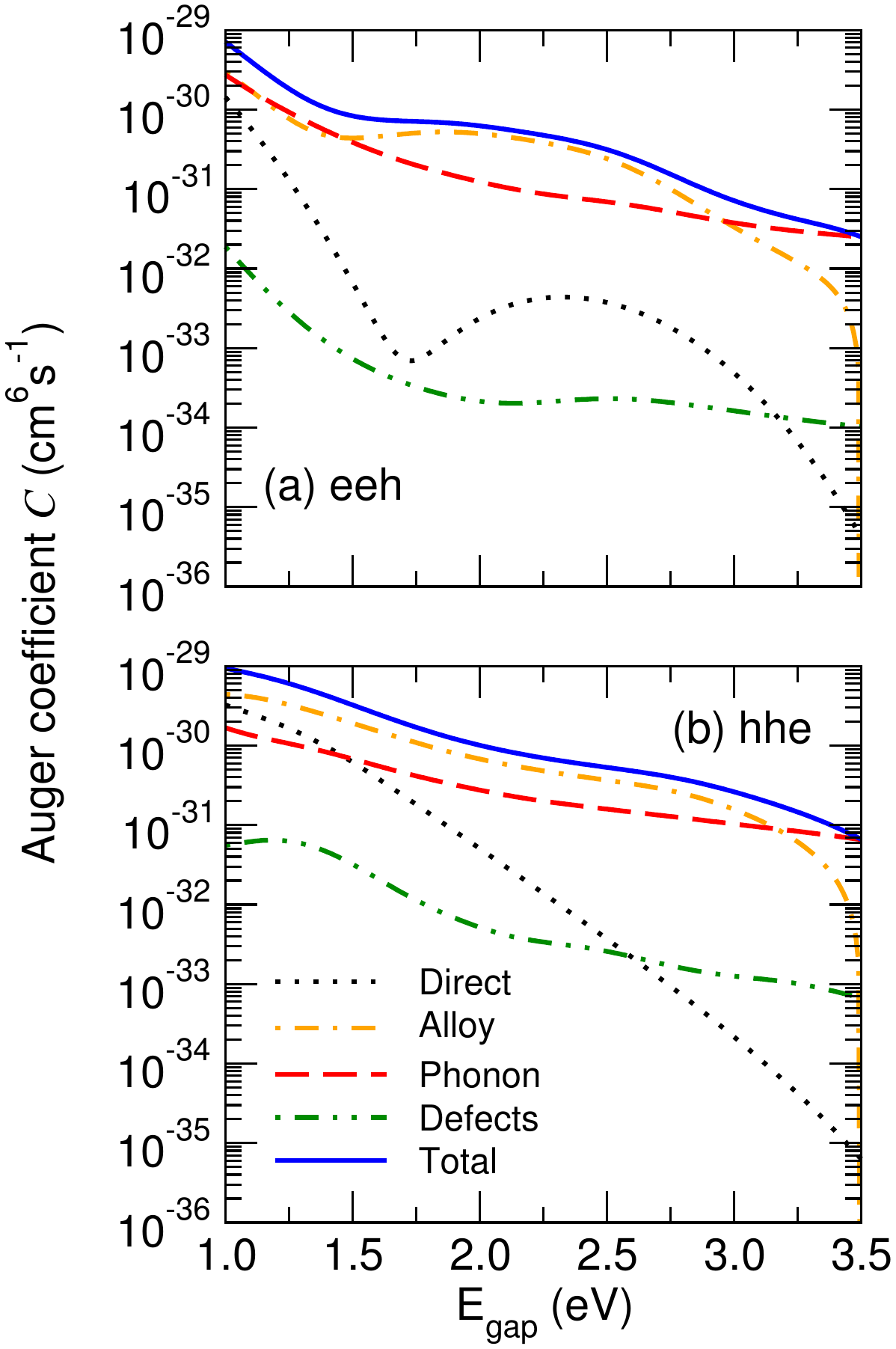}
\caption{\label{fig:auger_cumulative}
(Color online) Summary of the calculated results for the investigated direct and indirect Auger recombination mechanisms
in InGaN for free electron and hole concentrations of 10$^{19}$ cm$^{-3}$ and
a temperature of 300 K.
The contributions by the individual processes as well as the cumulative Auger coefficients are shown for (a) electron-electron-hole and (b) hole-hole-electron
processes.
In both cases, the results show that phonon-assisted Auger recombination is the dominant Auger recombination mechanism in InGaN.
}
\end{figure}

The contributions to the total Auger coefficients, as well as the cumulative result, for a carrier concentration of 10$^{19}$ cm$^{-3}$ as a function of the band gap
of InGaN are plotted in Fig.~\ref{fig:auger_cumulative}.
The band-gap dependence of the direct, phonon-assisted, and charged-defect-assisted
Auger coefficients is obtained by applying a rigid scissors shift to the band gap of GaN.
The alloy-scattering-assisted Auger coefficients calculated for In$_{0.25}$Ga$_{0.75}$N have been extrapolated to model other alloy compositions by
applying a rigid scissors shift to the band gap (which was the only effect included in Fig.~\ref{fig:ingan25_auger_vs_nk}) and multiplying with $x(1-x)/(0.25\cdot 0.75)$.
This is because the alloy-scattering-assisted Auger recombination rate is proportional to
the square of the alloy-disorder-scattering matrix elements, which are proportional to $x(1-x)$\cite{Ridley}.
The band-gap bowing equation of In$_x$Ga$_{1-x}$N from Ref.~\onlinecite{Moses2010} was inverted to yield the alloy composition as a function of the band gap.
Our results reveal that the primary contribution to Auger recombination in In$_x$Ga$_{1-x}$N alloys stems from
the scattering of charge carriers by the alloy disorder, while
phonon-mediated processes
provide a secondary contribution to Auger recombination.
Our results also show that the recombination rate due to direct Auger processes is negligible compared to the phonon- and alloy-scattering assisted terms, and that charge-defect-assisted Auger recombination is negligible for realistic defect densities in actual devices.
Our present results for the phonon-assisted and alloy-scattering-assisted coefficients slightly differ from values we reported previously,\cite{kioupakis:161107}
mainly due to the different model used in the present work to describe the screening by free carriers.

The calculated values for the Auger coefficients are in agreement with experimentally reported values for nitride LEDs and lasers,
and thus they highlight the important role of these non-radiative recombination mechanisms
during the operation of nitride optoelectronic devices at high power\cite{kioupakis:161107}.
The total value of the Auger coefficient for the e--e--h process ranges from
$7\times10^{-32}$ cm$^{6}$s$^{-1}$ for a gap of 3.0 eV to
$4\times 10^{-31}$ cm$^{6}$s$^{-1}$ when the gap is 2.4 eV, while the h--h--e auger coefficient ranges
from $3\times 10^{-31}$ cm$^{6}$s$^{-1}$ for a gap value of 3.0 eV to
$6\times 10^{-31}$ cm$^{6}$s$^{-1}$ for a 2.4 eV gap. The overall Auger coefficient accounting for both e--e--h and h--h--e
contributions amounts to $0.3$ -- $1 \times 10^{-30}$ cm$^{6}$s$^{-1}$ in the range of In
compositions corresponding to LEDs in the violet to green part of the visible spectrum.

\section{Summary}
In this work, we presented a first-principles methodology
and the associated computational details
for the determination of the direct and indirect Auger recombination
rate in direct-band-gap semiconductors.
With minor modifications the methodology can also be applied to indirect-gap materials.
The formalism has been applied to study Auger recombination in the group-III nitride materials used in optoelectronic devices.
The sensitivity to various parameters and approximations used in the calculations has been examined.
The results indicate that direct Auger recombination is not important in this class of wide-band-gap materials.
Indirect Auger processes, mediated by electron-phonon and alloy scattering,
are much stronger and the corresponding coefficients are
of the magnitude that has been shown to be sufficient to account for the efficiency droop of nitride LEDs\cite{kioupakis:161107}.
The formalism is not limited to nitrides and can be applied to study Auger recombination in other classes
of materials both from a fundamental materials physics point of view and to understand their performance
in high-power optoelectronic applications.
Future improvements to the methodology could include the first-principles calculation of the dielectric function within the random-phase approximation\cite{HybertsenLouie86} and the inclusion of alloy effects on the band structure and matrix elements for phonon-assisted Auger recombination.

\section{acknowledgments}
We thank Q. Yan, P. G. Moses, A. Janotti, J. Speck, and C. Weisbuch
for useful discussions.
E. K. acknowledges support by the National Science Foundation CAREER award through Grant No. DMR-1254314.
Work by D. S. and C. V.d.W.  was supported by the U.S. Department of Energy, Office of Science, BES, under Award No. DE-SC0010689.
P. R. acknowledges the support of the Deutsche Forschungsgemeinschaft, the UCSB-MPG Exchange Program and the NSF-IMI Program
(Grant No. DMR08-43934).
This research has been supported by the Academy of Finland through its Centres of Excellence Program (project no. 251748).
Computational resources were provided by the Center for Scientific Computing at the CNSI and MRL (an NSF MRSEC, DMR-1121053) (NSF CNS-0960316) and the DOE NERSC Facility supported by the DOE Office of Science (DE-AC02-05CH11231).
Figure~\ref{fig:ingan25_sqs_structure} was generated with the VESTA software.\cite{VESTA}


%

\end{document}